\documentclass[prd,onecolumn,amsmath,amssymb,floatfix,superscriptaddress,notitlepage]{revtex4-1}

\usepackage{graphicx}
\usepackage{amssymb}
\usepackage{amsmath}
\usepackage{hyperref}

\usepackage{color}

\DeclareMathAlphabet\mathbfcal{OMS}{cmsy}{b}{n}
\definecolor{darkgreen}{cmyk}{0.85,0.2,1.00,0.2} 
\definecolor{purple}{cmyk}{0.5,1.0,0,0} 

\definecolor{darkred}{cmyk}{0,1.00,1.00,0.3}

\DeclareTextSymbol{\degre}{T1}{23}

\newcommand{\mnras}{Mon.~Not.~Roy.~Astron.~Soc.}
\newcommand{\physrep}{Phys.~Rep.}

\newcommand{\apjl}{\apj~Lett.}
\newcommand{\aap}{Astronomy \& Astrophysics}

\newcommand{\vk} {\vec{k}}
\newcommand{\vq} {\vec{q}}
\newcommand{\vl} {\vec{l}}
\newcommand{\vx} {\vec{x}}
\newcommand{\vtheta} {\vec{\theta}}

\newcommand{\deltak} {\delta^K}
\newcommand{\deltad} {\delta_D}

\newcommand{\hdelta} {\hat{\delta}}
\newcommand{\on} {\bar{n}}

\newcommand{\rholin}{{\rho_\text{lin}}}
\newcommand{\hf} {\hat{f}}

\newcommand{\hn} {\hat{n}}

\newcommand{\beq} {\begin{equation}}
\newcommand{\eeq} {\end{equation}}
\newcommand{\bal} {\begin{aligned}}
\newcommand{\eal} {\end{aligned}}

\newcommand{\ave}[1]{\left\langle #1\right\rangle}

\newcommand{\mbf}[1]{\mbox{\boldmath$#1$}}

\begin{document}

\title[Cluster counts and bispectrum cross-covariance]
{Joint likelihood function of cluster counts and $n$-point correlation
functions: Improving their power through including halo sample variance}

\author{Emmanuel Schaan}
\affiliation{Department of Astrophysical Sciences, Princeton University, Peyton Hall, Princeton NJ 08544, USA}
\author{Masahiro Takada}
\affiliation{Kavli Institute for the Physics and Mathematics of the
Universe (WPI), Todai Institutes for Advanced Study, The University of Tokyo, Chiba 277-8582, Japan}
\author{David N. Spergel}
\affiliation{Department of Astrophysical Sciences, Princeton University, Peyton Hall, Princeton NJ 08544, USA}
\affiliation{Kavli Institute for the Physics and Mathematics of the
Universe (WPI), Todai Institutes for Advanced Study, The University of Tokyo, Chiba 277-8582, Japan}

\begin{abstract} 
Naive estimates of the statistics of large scale structure and weak lensing power spectrum measurements that include only Gaussian
errors exaggerate their scientific impact.   Non-linear evolution and finite volume effects are both significant sources
of non-Gaussian covariance that reduce the ability of power spectrum measurements to constrain cosmological parameters.
Using a halo model formalism, we derive 
an
intuitive understanding of the various contributions to the covariance and show that  our analytical treatment agrees 
with simulations.  This approach enables an approximate derivation of 
a joint likelihood for the cluster
number counts, the weak lensing power spectrum and the bispectrum.  We show that this likelihood is
a good description of the ray-tracing simulation.  Since all of these observables are sensitive to the same finite volume effects
and contain information about the non-linear evolution, a combined analysis recovers much of the ``lost" information and
obviates the non-Gaussian covariance.  For upcoming weak lensing surveys, we estimate that a joint analysis of power spectrum, number
counts and bispectrum will produce
an improvement of about $30-40\%$ in determinations of the matter density and the scalar amplitude.
This improvement is
equivalent to doubling the survey area.
\end{abstract}

\maketitle

\section{Introduction}
\label{sec:intro}

Understanding the nature of dark energy is the aim of many ongoing and 
upcoming galaxy
surveys such as 
the Baryon Oscillation Spectroscopic Survey
(BOSS)\footnote{\url{http://cosmology.lbl.gov/BOSS/}},
the Kilo-Degree Survey
(KiDS)\footnote{\url{http://www.astro-wise.org/projects/KIDS/}},
the Extended BOSS (eBOSS)\footnote{\url{http://www.sdss3.org/future/eboss.php}},
the Dark Energy Survey
(DES)\footnote{\url{http://www.darkenergysurvey.org}} \cite{DES},
the Subaru Hyper Suprime-Cam (HSC) survey\footnote{\url{http://www.naoj.org/Projects/HSC/index.html}}
\cite{Miyazakietal:06}, 
the Subaru Prime Focus Spectrograph
(PFS)\footnote{\url{http://sumire.ipmu.jp/en/2652}}\cite{Takadaetal:12},
the Dark Energy Spectroscopic Instrument (DESI)\footnote{\url{http://desi.lbl.gov}}, 
the Large Synoptic Survey Telescope (LSST)
\cite{LSSTScienceBook}, 
the ESA satellite mission Euclid \cite{EuclidDefinitionStudyReport},
 and
the NASA satellite mission WFIRST \cite{WFIRST}.

The science yield from these surveys appears to be less than one would
naively expect. If we were observing a density field in the linear
regime, the various modes would be uncorrelated and the amount of
information available would scale with
the number of modes.  However, 
for weak
lensing, the scales of interest ($l\sim 10^3$)
 are well into the non-linear regime \cite{JainSeljak:97,Jainetal:00}.
On such scales, mode couplings induce non-Gaussian features, which move
information from the power spectrum to the bispectrum and higher
$n$-point correlation functions, and induce extra correlated scatter on the various
$n$-point 
functions,
as well as on their various multipoles
\cite{Scoccimarroetal:99,HuWhite:01,CoorayHu:01,TakadaJain:04,TakadaJain:09}.  In order to recover the information
diluted between the various multipoles and $n$-point correlation functions, one has
to combine them in a joint analysis, and doing so requires an
understanding of their non-Gaussian correlated errors
\cite{TakadaJain:04,Kayoetal:12,KayoTakada:13}.

Another important source of correlated scatter comes from a
finite-volume effect
of the survey domain: density modes with wavelengths larger than
the volume are not measurable from within the survey, but affect the
observables in a predictable way
\cite{Hamiltonetal:06,HuKravtsov:03,Sefusattietal:06,TakadaBridle:07,TakadaJain:09,Satoetal:09,Kayoetal:12,TakadaHu:13,TakadaSpergel:13,Lietal:14}. 
This is another reason to combine
various observables on the same survey: since they are all affected by
the same long wavelength modes, a combined analysis can determine the amplitude of  these
non-directly observable modes \cite{TakadaBridle:07,Kayoetal:12,TakadaSpergel:13}.

In other words, recovering the non-Gaussian information and calibrating
out the long wavelength modes inaccessible from within the survey are
two reasons to combine probes, and therefore justify the need for
understanding their covariances. Although the first effect has been
understood for a long time, progress on the second one is only
recent. In Ref.~\cite{TakadaHu:13}, this `super-sample covariance' 
in power spectrum measurement
is
described in terms of mode coupling through the window function and a
`trispectrum consistency relation'. In Ref.~\cite{TakadaSpergel:13}, this
`halo sample variance' is a consequence of the halo model: an upscatter
in the average density triggers an upscatter in halo counts through
linear biasing, which leads to a coherent upscatter in the various
$n$-point correlation functions.

Our results derive straightforwardly from the same assumptions as the
halo model: the matter overdensity field can be expressed by the
distribution of halos in different mass bins,
and the halos form a biased Poisson
sampling of the underlying density field.
The halo sample variance should therefore be
considered a standard prediction of the halo model, just as much as
usual decomposition $P = P^{1h} + P^{2h}$ for the power spectrum.

Our study builds on Ref.~\cite{TakadaSpergel:13} and generalizes the
results therein to all $n$-point correlation functions. The starting
point of our analysis is the expression of the matter overdensity in
terms of halos, instead of the decomposition $P = P^{1h} + P^{2h}$,
which allows for a consistent derivation of the auto- and
cross-covariances between halo number counts and $n$-point functions.

The outline of this paper is the following. In Section~\ref{sec:method},
we present our assumptions and derive the general formula for the halo
sample variance contributions. We apply these results to the cluster
counts and the matter $n$-point correlation functions in Section~\ref{sec:3d}. Then
we apply the formulation to the 2D fields, the angular number counts of
clusters and the 
lensing convergence $n$-point correlation functions in Section~\ref{sec:2d}, which
allows us to check our results against ray-tracing simulations from
Ref.~\cite{Satoetal:09}. In Section~\ref{sec:likelihood_fisher}, we present
an approximate joint likelihood for the cluster counts and lensing
convergence power spectrum and bispectrum. We compare it to simulations
and use it to forecast an estimation of cosmological parameters obtained 
when combining the cluster counts to the lensing power spectrum and
bispectrum for a future galaxy survey.

\section{Method}
\label{sec:method}

In this section, we briefly review the halo model ingredients
 that we
shall use to derive the halo sample variance. 
We present our notations and method to take into account the
effects of a finite-volume survey, and give the general derivation for
the halo sample variance for any observable.

\subsection{Standard halo model}

The halo model
\cite{Seljak:00,PeacockSmith:00,MaFry:00,CooraySheth:02,HuKravtsov:03}
is based on the assumption that all matter in the universe is contained in 
halos of some mass scale. The mass function 
$dn/dm$
gives the mean number density for halos of mass $m$, and the density
profile of halos $u_m(\vx)$, defined so as to satisfy the normalization condition 
$\int\!{\rm d}^3\vx \, u_m(\vx)=1$,
is assumed to depend only on their mass $m$, at a given redshift. 
Thus the halo model expresses the observed matter overdensity as the familiar sum over halos:
\beq
\hdelta(\vk) = \int dm \left( \frac{m}{\bar{\rho}} \right) u_m(\vk)
\frac{dn}{dm}\hdelta^h_m(\vk),
\label{eq:deltak}
\eeq
where $\hdelta^h_m(\vk)$ is the number density fluctuation field
(its
Fourier transform) for halos of mass $m$.
In what follows, we shall replace the integral with a sum over mass bins
 $i$, with mean mass $m_i$ and bin width $\Delta m_i$:
\beq
\label{eq:def_hdelta}
\hdelta (\vk) = \sum_i \left( \frac{m_i}{\bar{\rho}} \right) u_i(\vk) \, \hn_i(\vk, \delta_b).
\eeq
Here $\hn_i(\vk)$ is the halo number density field,
defined as
\beq
\hn_i(\vx) = \left.\frac{dn}{dm}\right|_{m_i}
\Delta m_i \left[  1 + \hdelta^h_i(\vx)  \right].
\eeq
The observed number of halos in the $i$-th mass bin $m_i$, for
a given small volume $\delta V$ around the position $\vx$, is simply
given as $\hn_i(\vx)\delta V$.
The halo model formulation above is useful for our purpose.
First, Eq.~\eqref{eq:def_hdelta} shows that the statistical properties
of the matter 
density field
$\hdelta (\vk)$ are 
determined by the halo number density field
$\hn_i(\vk)$ and the halo mass profile $u_i(\vk)$. 
Second, 
Eq.~\eqref{eq:def_hdelta} allows us to straightforwardly compute
cross-correlations
between matter $n$-point functions and 
the number counts of halos as we will show below.

We assume that 
 the halo number density in a volume element $\delta V$ around the
 position $\vx$
follows a Poisson statistics, with mean determined by the underlying density field $\rholin(\vx)$:
\beq
\bal
\label{eq:poisson}
\ave{ \hn_i(\vx_1) \hn_j(\vx_2) }_{\text{Pois.} | \rholin }
&=
n_i(\vx_1) n_j(\vx_2)
+ \deltak_{ij} \deltad(\vx_1 -\vx_2) n_i(\vx),
\eal
\eeq
where $n_i(\vx)$ is the mean halo number density for the volume
$\delta V$ around $\vx$, for a
fixed 
$\rholin(\vx)$, defined 
as 
$n_i(\vx) \equiv \ave{ \hn_i(\vx) }_{ \text{Pois.} | \rholin }$, 
and $\delta^K_{ij}$ is the Kronecker delta function:
$\delta^K_{ij}=1$ if $i=j$, otherwise $\delta^K_{ij}=0$.
We
have assumed that the halo number densities of different mass bins
are independent.
We assume that 
the halo number densities are given as biased tracers of the linear
density field $\rholin(\vx)$:
\beq
\label{eq:bias}
n_i(\vk) = \on_i \, b_i \, \delta_\text{lin}(\vk)
= \left.\frac{dn}{dm}\right|_{m_i} \Delta m_i \, b_i \, \delta_\text{lin}(\vk),
\eeq
were $\on_i \equiv \ave{ n_i }_\rholin \equiv
 \ave{ \hn_i(\vx) }_{ \text{Pois.}, \rholin } =
\left.dn/dm
\right|_{m_i} \Delta m_i $ is the 
ensemble average of the
halo number
 density, obtained by
marginalizing over Poisson sampling and different realizations of 
the linear density field $\rholin(\vx)$, and $b_i$ is the linear
bias for halos of the $i$-th mass bin, $b_i\equiv b(m_i)$.

Using the matter density field
$\hdelta(\vk)$ in Eq.~\eqref{eq:def_hdelta}, and the properties \eqref{eq:poisson} and \eqref{eq:bias},
it is straightforward
to express the matter power spectrum  $\ave{ \hdelta(\vk) \hdelta(\vk')}$ in terms of the halo number counts:
\beq
\label{eq:ex1}
\bal
\ave{ \hdelta(\vk) \hdelta(\vk ')}
&= 
\left(2\pi\right)^3 \deltad(\vk+\vk')
\left[
\sum_i \left( \frac{m_i}{\bar{\rho}} \right)^2
\on_i |u_i(k)|^2
+\sum_{i,j}  \left( \frac{m_i m_j}{\bar{\rho}^2} \right) 
\on_i\on_j 
u_i(k)u_j(k) \,\,
b_ib_j P_\text{lin} (k) \right],
 \eal
\eeq
where $P_\text{lin}$ is the linear matter power spectrum.
The first term is the 1-halo term, $P^{1h}(k)$, 
arising from correlations between
matter in the same halo, while the second term is the 2-halo term, 
$P^{2h}(k)$, arising from matter in two different halos.

\subsection{Finite survey effect: method}

In this paper, we study the finite-volume effect of a survey on
$n$-point correlation function measurements. We 
characterize
the survey 
by its three-dimensional volume, $V_S$, and 
the average density
contrast across the survey region, 
$\delta_b$. The 
super-survey mode 
is defined as
$\delta_b\equiv\int\!d^3\vx \, W^{3D}(\vx)\hdelta(\vx)$, 
where $W^{3D}(\vx)$ is the survey window function; $W^{3D}(\vx)=1/V_S$ if 
$\vx$ is in the survey region, otherwise $W^{3D}(\vx)=0$. 
Note that the window function satisfies the normalization condition 
$\int\!d^3\vx\, W^{3D}(\vx) = 1$.
For simplicity, we
neglect effects of gradients or tidal fields of the super-survey
density field as well as an effect of incomplete selection or weights.

In the presence of the super-survey mode $\delta_b$, the
expectation value of the halo number density 
is
biased compared to the ensemble average:
\beq
\label{eq:linear_bias}
\on_i(\delta_b) = \ave{ \hn_i(\vx) }_{\text{Pois.}, \rholin | \delta_b}
= \on_i \left[  1 + b_i \delta_b  \right].
\eeq
For a sufficiently large survey volume,
$\delta_b$ 
can be safely considered to be 
in the linear regime and  
obey 
a Gaussian 
distribution
with variance 
\begin{equation}
\sigma_m^2(V_S) \equiv \int \frac{d^3\vk}{(2\pi)^3} | W^{3D}(\vk)
|^2 P_\text{lin} (k).
\label{eq:sigmam} 
\end{equation}
Thus, marginalizing over realizations of the super-survey mode $\delta_b$ as in Refs.~\cite{HuKravtsov:03,HuCohn:06} gives:
\beq
\label{eq:substitution_n}
\bal
&\ave{\on_i(\delta_b) }_{\delta_b} = \ave{ \hn_i(\vx) }_{\text{Pois.}, \rholin} = \on_i \\
&\ave{ \bar{n}_{i_1}(\delta_b) ... \bar{n}_{i_N}(\delta_b) }_{\delta_b}
=
\bar{n}_{i_1} ... \bar{n}_{i_N} \left[  1 + \sigma_m^2(V_s) \sum_{ \{ j,l\} \in \{1, ..., N\} } b_{i_j} b_{i_k}  \right] .
\eal
\eeq

Combining Eq.~(\ref{eq:ex1}) with Eq.~(\ref{eq:linear_bias}), 
we can express the power
spectrum estimator, drawn from the same finite-volume survey region,
 in terms of the halo number density fluctuations as
\begin{eqnarray}
\ave{ \hdelta(\vk) \hdelta(\vk ')}_{| \delta_b}
&=& 
\left(2\pi\right)^3 \deltad(\vk+\vk')
\left[
\sum_i \left( \frac{m_i}{\bar{\rho}} \right)^2
\on_i \left[  1 + b_i \delta_b  \right]
|u_i(k)|^2
\right.\nonumber\\
&&\left. 
+\sum_{i,j}  \left( \frac{m_i m_j}{\bar{\rho}^2} \right) 
\on_i\on_j \left[  1 + b_i \delta_b  \right] \left[  1 + b_j\delta_b  \right]
u_i(k)u_j(k) \,\,
b_ib_j P_\text{lin} (k) \right].
\end{eqnarray}
We then marginalize over the Gaussian variable $\delta_b$, using Eq.~(\ref{eq:substitution_n}), to
obtain the expectation value of power spectrum estimator:
\beq
\label{eq:ps_halo1}
\bal
\ave{ \hdelta(\vk) \hdelta(\vk ')}
&= 
\left(2\pi\right)^3 \deltad(\vk+\vk')
\left[
\sum_i \left( \frac{m_i}{\bar{\rho}} \right)^2
\on_i
|u_i(k)|^2
+\sum_{i,j}  \left( \frac{m_i m_j}{\bar{\rho}^2} \right) 
\on_i\on_j \left[  1 + b_i b_j  \sigma_m^2(V_S) \right]
u_i(k)u_j(k) \,\,
b_ib_j P_\text{lin} (k) \right].
 \eal
\eeq
The 1-halo term (first term) is unchanged from Eq.~\eqref{eq:ex1}, whereas
the 2-halo term (second term) gets a correction term proportional to $b_i b_j
\sigma_m^2$, due to the finite volume effect. Since $b_i b_j  \sigma_m^2
\ll 1$ for a large survey volume of interest, 
this correction is safely
negligible. 
Hence the mean value of our power spectrum estimator is unchanged. However, as we shall see in the next section, its covariance is affected by the finite volume of the survey.

\subsection{Finite survey effect: general derivation}
\label{subsec:general_derivation}

In this section we give a general discussion on the effect of
a
finite-volume survey 
on
observables.
Consider observables $\hat{f}$ and $\hat{g}$ that probe the matter
density fluctuation field
through
the halo number density field, 
$\left\{ \hn_i (\vx) \right\}$. 
For instance, 
$\hat{f}$ is the
number counts of halos in the $i$-th mass bin, 
$\hat{N}_i$,
or
the matter $n$-point function $\hat{P}_n$.
In the following,
we derive general expressions for the expectation value of $\hat{f}$
as well as
the co- or cross-variances between $\hat{f}$ or/and $\hat{g}$.

Suppose that $\hat{f}$ is an estimator of some observable and
that $\bar{f}(\delta_b) = \ave{\hat{f}}_{\text{Poisson}, \rholin |
\delta_b}$ 
is the expectation value  for survey 
realizations with a
{\it fixed} super-survey mode $\delta_b$.
Since the estimator $\hat{f}$ depends on
the halo number density field
$\left\{ \hn_i (\vx) \right\}$, 
the expectation value $\bar{f}(\delta_b)$ 
depends on
$\left\{ \on_i (\delta_b) \right\}$, which is the
expectation value of the halo number densities for realizations with
fixed $\delta_b$ (see Eq.~\ref{eq:linear_bias}). 
Therefore,
if marginalizing
$\bar{f}(\delta_b)$ over the Gaussian variable
$\delta_b$,
one can find that $\bar{f}(\delta_b)$ is now given as a function
of the covariances of the halo number densities such as 
$\ave{ \bar{n}_{i_1}(\delta_b)
... \bar{n}_{i_N}(\delta_b) }_{\delta_b}$ (see 
Eq.~\ref{eq:substitution_n}).
Then, as we have seen for the power spectrum case in
Eq.~(\ref{eq:ps_halo1}), $\bar{f}(\delta_b)$ generally has correction terms
proportional to $\sigma_m^2(V_s) \sum_{ \{ j,l\} \in \{1,
..., N\} } b_{i_j} b_{i_l}$.
However, the correction terms are
negligible as $\sigma_m^2(V_s) \ll
1$ for the cases of interest.
Hence the expectation value of the estimator $\hf$ is unaffected by the finite volume of the survey.

The situation is different for the covariance calculation.
Since $\bar{f}(\delta_b)$ is given as a
function of the $\left\{ \on_i (\delta_b) \right\}$ and
$|\delta_b|\ll 1$, 
we can Taylor expand 
$\bar{f}(\delta_b)$ 
 as
\begin{eqnarray}
 \bar{f}(\delta_b) &\simeq& \bar{f}(0) + 
\left.\frac{\partial \bar{f}}{\partial \delta_b}\right|_{\delta_b=0}
\delta_b + 
\frac{1}{2}\left.\frac{\partial^2\bar{f}}{\partial
\delta_b^2}\right|_{\delta_b=0}
\delta_b^2 + \mathcal{O}(\delta_b^3) \nonumber\\
&=&\bar{f}(0) + 
\sum_i\left.\frac{\partial \bar{f}}{\partial \bar{n}_i}
\frac{\partial \bar{n}_i}{\partial \delta_b}
\right|_{\delta_b=0}
\delta_b + 
\frac{1}{2}\sum_{i,j}
\left.\frac{\partial^2\bar{f}}{\partial \bar{n}_i
\partial \bar{n}_j}\frac{\partial \bar{n}_i}{\partial \delta_b}
\frac{\partial \bar{n}_j}{\partial \delta_b}
\right|_{\delta_b=0}
\delta_b^2 + \mathcal{O}(\delta_b^3) \nonumber\\
&=&\bar{f}(0) + 
\sum_i\left.\frac{\partial \bar{f}}{\partial \ln \bar{n}_i}
\right|_{\delta_b=0}b_i
\delta_b + 
\frac{1}{2}\sum_{i,j}
\left.\frac{\partial^2\bar{f}}{\partial \bar{n}_i
\partial \bar{n}_j}
\right|_{\delta_b=0}
\bar{n}_i \bar{n}_j 
b_ib_j\delta_b^2 + \mathcal{O}(\delta_b^3),
\end{eqnarray}
where the derivative such as $\partial \bar{f}/\partial \delta_b$
is with respect to $\delta_b$ with all other parameters being kept
fixed. Note that, in the third line on the r.h.s., we used
Eq.~(\ref{eq:linear_bias}) to obtain $\partial \bar{n}_i/\partial
\delta_b|_{\delta_b=0}=b_i\bar{n}_i$, and 
ignored a contribution from nonlinear halo bias,
i.e., set $\partial^2\bar{n}_i/\partial\delta_b^2|_{\delta_b=0}=0$ for
simplicity. 
A similar equation holds for any other observable, say $\hat{g}$.
After some straightforward algebra (see Appendix~\ref{app:pp}), we
can find that the cross-covariance between the two observables,
$\hat{f}$ and $\hat{g}$, is generally given as
\beq
\label{eq:general_exp}
\text{Cov}\left[  \hat{f}, \hat{g}  \right]
\simeq
\left\langle   \text{Cov}\left[  \hat{f}, \hat{g}  \right]_{\text{Pois.}, \rholin | \delta_b}   \right\rangle_{\delta_b} 
+ \sigma_m^2(V_S)  \frac{\partial \bar{f}}{\partial \delta_b}
\frac{\partial \bar{g}}{\partial \delta_b}
+ \mathcal{O} (\sigma_m^4) .
\eeq
The term
$\left\langle \text{Cov}\left[  \hat{f}, \hat{g}  \right]_{\text{Pois.},
\rho_\text{lin} | \delta_b} \right\rangle_{\delta_b}$ is 
the standard covariance term (with correction terms
such as a term proportional to $\sigma_m^2(V_S)b_ib_j$, 
but such terms are negligible in practice as we discussed above).
On the other hand, despite the small factor $\sigma_m^2$, the second
term gives a significant or even dominant contribution to the covariance
for a large-volume survey, as we shall see below. We hereafter call this
term the ``halo sample variance'' (HSV) term
\citep{HuKravtsov:03,Satoetal:09,TakadaSpergel:13,TakadaHu:13}.
In the following, 
we approximate the full covariance by a sum of the standard
covariance and the HSV term:
\beq
\label{eq:simplified_exp}
\text{Cov}\left[  \hat{f}, \hat{g}  \right]
\simeq
\text{Cov}\left[  \hat{f}, \hat{g}  \right]_{ \text{Pois.}, \rholin | \delta_b=0} 
+ \sigma_m^2(V_S)  \frac{\partial \bar{f}}{\partial \delta_b}
\frac{\partial \bar{g}}{\partial \delta_b}.
\eeq
The above derivation is similar to what is done in
Ref.~\cite{TakadaHu:13}, however is different in a sense that we derived
the HSV terms by fully relying on the setting and assumptions built into
the halo model approach.

\section{Halo number counts and $n$-point functions of the matter overdensity in 3d}
\label{sec:3d}

In this section, using the formulation in the preceding section (in particular Eq.~\ref{eq:simplified_exp}),
we compute the covariances of the halo number counts and the $n$-point
correlation functions as well as their cross-covariances.

\subsection{Halo number counts}

We assume that the number counts of halos in the $i$-th mass bin
can be estimated from a survey volume:
$\hat{N}_i = \int_{V_S}\! d^3\vx \,\, \hn_i(\vx, \delta_b)$. 
The ensemble average of the
number counts is simply
\begin{equation}
\ave{\hat{N}_i} = \bar{N}_i = \bar{n}_i V_S =\left.
 \frac{dn}{dm}\right|_{m_i} \Delta m_i V_S.
\end{equation}

Assuming
linear halo bias as in Eq.~\eqref{eq:linear_bias}, we can compute the first
derivative of $\bar{N}_i(\delta_b)$ with respect
to $\delta_b$:
$\partial \bar{N}_i/\partial \delta_b = b_i \bar{N}_i $.
Hence, from Eq.~\eqref{eq:general_exp}, we find the covariance
of the number counts to be
\begin{equation}
\label{eq:number_counts}
 \text{Cov}\left[  \hat{N}_i, \hat{N}_j  \right] = \deltak_{i,j}  \bar{N}_i + \sigma_m^2(V_S) b_i b_j \bar{N}_i \bar{N}_j.
\end{equation}
The halo number counts of different mass bins thus become
correlated with each other through the super-survey mode
$\delta_b$, as found in Refs.~\cite{HuKravtsov:03,HuCohn:06,TakadaBridle:07,TakadaSpergel:13}.
Ref.~\cite{Crocceetal:10} showed that 
the above covariance well reproduces the simulation results, while the
theory underestimates the simulation, if including the first term alone,
i.e., the Poisson error assumption.

\subsection{Covariances of $n$-point matter correlation functions}

Now let us consider covariances of the $n$-point matter
correlation functions. The underlying true 
$n$-point correlation function, 
$\bar{P}_n(\vk_1,\dots,\vk_n)$, is defined as
\beq
\label{eq:def_p}
\ave{\delta(\vk_1) ... \delta(\vk_n)}_c =  V_S 
\deltak_{\vk_1+...+\vk_n}   \bar{P}_n(\vk_1, ..., \vk_n),
\eeq
where here $\delta(\vk)$ refers to the true matter overdensity, as
opposed to the one observed from a finite box, and we substituted $V_S
\deltak_{\vk_1+...+\vk_N} $ to the usual $\left( 2\pi \right)^3
\deltad(\vk_1+...+\vk_N)$, as appropriate when using discrete Fourier
transform \citep{TakadaBridle:07,Kayoetal:12}.

\subsubsection{Power spectrum}

For a finite-volume survey,
we define an estimator of the power spectrum as
\beq
\label{eq:p_estimator}
\hat{P}(k) \equiv  \frac{1}{N(k)V_S} \sum_{|\vq| \simeq k} 
 \hdelta(\vq) \hdelta(-\vq),
\eeq
where the average is over a shell of wavevectors $\vq$ which have
lengths of $k$, with a shell width $\Delta k$, and $N(k)$ is the number of
independent Fourier modes in the shell, approximated as 
$N(k) \simeq k^2 \Delta k V_S/(2 \pi^2)$ for the limit 
$k\gg 1/V_S^{1/3}$.

The ensemble average of the estimator (Eq.~\ref{eq:p_estimator}) gives
the underlying true power spectrum, with a negligible, small correction
as we discussed in  Section~\ref{sec:method}. Employing the halo model
approach, we can derive the ensemble-average power spectrum
\citep[Ref.][also see Appendix~\ref{app:pp} for the detailed derivation]{TakadaSpergel:13}:
\beq
\label{eq:p1h2h}
\bar{P}(k) = \sum_i \bar{n}_i p_i^{1h}(k)
+
\sum_{i,j} \bar{n}_i \bar{n}_j p_{ij}^{2h}(k),
\eeq
where 
$p_i^{1h}(k)  \equiv (m_i/\bar{\rho})^2|u_i(k)|^2$
and $p_{ij}^{2h}(k) \equiv (m_i m_j/\bar{\rho}^2) 
u_i(k) u_j(k) b_ib_j P_\text{lin} (k)$.
For the above halo model expression, we discretized the mass function
integrals 
into a summation over halo mass bins.

Inserting the power spectrum estimator into
Eq.~(\ref{eq:simplified_exp}), we can derive an expression of the power
spectrum covariance including the HSV effect
(see Appendix~\ref{app:pp} for the detailed derivation):
\begin{eqnarray}
 \label{eq:covp}
\text{Cov}\left[  \hat{P}(k), \hat{P}(k')  \right]
&=&
\frac{2 \delta^K_{k,k'} } {N(k)}   \bar{P}^2(k)
+\frac{1}{V_S} 
\bar{T}(k,k')
\nonumber \\
&&\hspace{-3em}
+ \sigma_m^2(V_S)
\left[ \int dm \frac{d\on}{dm}  b(m) \left( \frac{m}{\bar{\rho}} \right)^2 | u_m(k) |^2 
+   
2\left( \int dm \frac{d\on}{dm}  b^2(m) \left( \frac{m}{\bar{\rho}}
					\right) u_m(k)
 \right)\right.\nonumber\\
&&\hspace{-1em}\left.\times
\left( \int dm' \frac{d\on}{dm}  b(m') \left( \frac{m'}{\bar{\rho}} \right) u_{m'}(k) \right)
P_\text{lin} (k)
\right] 
\times \left[\frac{ }{ } k \leftrightarrow k'\right],
\end{eqnarray}
where $\bar{T}(k,k')$ is the angle-averaged squeezed trispectrum.  The first
and second terms on the r.h.s. are the standard Gaussian and
non-Gaussian terms \cite{Scoccimarroetal:09}. The former contributes
only to diagonal elements of the covariance matrix, while the latter
describes correlations between the power spectra of different modes
arising from the connected 4-point correlation function.  Both terms
scale with the survey volume as $1/V_S$.

The third term is the HSV term arising from
correlations of Fourier modes inside the survey volume with super-survey
modes \cite{Satoetal:09,Kayoetal:12,TakadaHu:13}. 
The HSV depends on the rms density fluctuations
of the survey volume, $\sigma_m^2(V_S)$ (Eq.~\ref{eq:sigmam}). As discussed
in Section~\ref{sec:method}, the HSV terms are given in
terms of the response of the power spectrum to the super-survey mode
$\delta_b$; $\sigma_m^2(V_S) (\partial \bar{P}(k)/\partial \delta_b)
(\partial \bar{P}(k')/\partial \delta_b)$,
where we used the halo model to compute the derivatives.  
The HSV contribution
in Eq.~(\ref{eq:covp}) includes the response of
the 1-halo term, that of the 2-halo term and their cross terms. 
Physically this effect can be interpreted as follows: 
if the survey volume is embedded
in an overdensity region,  $\delta_b>0$, it increases
the halo number counts, and then causes an up-scatter in the power
spectrum estimate coherently over different
$k$-bins.  
The HSV terms depend on the survey volume via $\sigma^2_m(V_S)$
(Eq.~\ref{eq:sigmam}), which generally has a different dependence from
$1/V_S$.

\subsubsection{Bispectrum}

Similarly to the power spectrum, we can define the bispectrum
estimator
for a triangle configuration that is specified by three side lengths
($k_1,k_2,k_3$):
\beq
\hat{B}(k_1, k_2, k_3) \equiv  \frac{1}{N_\Delta(k_1, k_2, k_3) V_S} 
\sum_{\vq_i; q_i\in k_i
}
\deltak_{\vq_1 + \vq_2 + \vq_3}
\hdelta(\vq_1) \hdelta(\vq_2) \hdelta(\vq_3),
\eeq
where the summation runs over all the triplets of 
the Fourier field, $\{\hdelta(\vq_i)\}$,
that
form the triangle configuration within the bin widths, and the Kronecker
delta function $\delta^K_{\vq_1+\vq_2+\vq_3}$ imposes the triangle
configuration condition in Fourier space. 
The quantity $N_\Delta(k_1,k_2,k_3)$ is the
number of independent triplets for the triangle configuration, defined as
\beq
N_\Delta(k_1, k_2, k_3) \equiv 
\sum_{\vq_i; q_i\in k_i
}
\deltak_{\vq_1 + \vq_2 + \vq_3} .
\eeq

Within the halo model framework 
the ensemble average of the bispectrum estimator is given by the sum of
the 1-, 2- and 3-halo terms as
\beq
\bar{B} = 
\sum_i \bar{n}_i b_i^{1h}
+\sum_{i,j} \bar{n}_i \bar{n}_j b_{ij}^{2h}
+ \sum_{i,j,l} \bar{n}_i \bar{n}_j \bar{n}_l b_{ijl}^{3h} ,
\eeq
where the summation of each term runs over halo mass bins.

Ref.~\cite{Kayoetal:12} derived the bispectrum covariance
including the HSV terms. In Appendix~\ref{app:pp}, we
revisit
the covariance derivation under our formulation, where we
include the HSV contributions  to the 1-, 2- and
3-halo terms by computing the response to the super-survey
modes, $\partial \bar{B}/\partial \delta_b$.
Contrary to the case of the power spectrum, 
we find that the HSV effects in 
the 2- and 3-halo terms
are negligible,
and therefore we consider the 1-halo term alone for the HSV effect in the
following:
\begin{eqnarray}
\label{eq:hsv_b}
{\rm Cov}[B(k_1,k_2,k_3),B(k^\prime_1,k^\prime_2,k^\prime_3)]^{\rm HSV}
&=&
\sigma_m^2(V_S)
\left[\int dm \frac{d\on}{dm} b(m) \left( \frac{m}{\bar{\rho}} \right)^3 u_m(k_1) u_m(k_2) u_m(k_3) \right]\times \left[\frac{}{} k \leftrightarrow k'\right] .
\end{eqnarray}

\subsubsection{Covariance between $n$- and $n'$-point correlation functions}
\label{sec:n-point}

Similarly, we can estimate the cross-covariance 
between the $n$- and $n'$-point correlation functions. For instance, 
the HSV
term in the cross-covariance
between power spectrum and bispectrum can be computed from the response
involving $\sigma_m^2(V_S) (\partial P/\partial \delta_b)(\partial B/\partial
\delta_b)$.

\subsection{Cross-correlation between $n$-point functions and halo number counts}
\label{sec:ccov3D}

When the halo number counts and the matter $n$-point correlation
function are drawn from the same survey region, the two are correlated
with each other, because both probe the underlying matter density field
in large-scale structure.

In the case of the power spectrum,
Eq.~\eqref{eq:simplified_exp}
leads to
\beq
\bal
&\text{Cov}\left[  \hat{N}_i, \hat{P}(k)  \right]
=
 \bar{n}_i 
\left[ p_i^{1h} (k) + 2\sum_j \bar{n}_j p_{ij}^{2h} (k) \right]
+
\sigma_m^2(V_S) b_i \bar{N}_i 
\left[
\sum_j b_j \bar{n}_j p_j^{1h}(k)
+
2 \sum_{j,l} b_j \bar{n}_j \bar{n}_l p_{jl}^{2h}(k)
\right] .
\eal
\eeq
For the bispectrum case, the cross-covariance is
\begin{eqnarray}
\text{Cov}\left[  \hat{N}_i, \hat{B} \right]
&=& 
 \bar{n}_i 
\left[ 
b_i^{1h}
+
2 \sum_j \bar{n}_j b_{ij}^{2h}
+
3 \sum_{j,l} \bar{n}_j \bar{n}_l b_{ijl}^{3h}
\right] 
\nonumber \\
&&+\sigma_m^2(V_S) b_i \bar{N}_i 
\left[
\sum_j b_j \bar{n}_j b_j^{1h} 
+
2 \sum_{j,l} b_j \bar{n}_j \bar{n}_l b_{jl}^{2h} 
+
 3 \sum_{j,l,m} b_j \bar{n}_j \bar{n}_l \bar{n}_m b_{jlm}^{3h}
\right] .
\end{eqnarray}

In both cases, the terms in the first square bracket on the
r.h.s. arise from Fourier modes inside the survey volume due to the
Poisson nature of the halo number counts, and correspond to $\overline{n}_i \left(\partial \bar{P}_n / \partial \overline{n}_i \right)$ .
The terms in the second square bracket are the HSV terms, and correspond to $\sigma_m^2(V_S) (\partial \bar{N}_i /\partial \delta_b)(\partial \bar{P}_n /\partial
\delta_b)$.
Thus the super-survey mode
$\delta_b$ causes a co-variance in the number counts and the $n$-point
correlation functions.

\section{Application to lensing convergence and clusters number counts} 
\label{sec:2d}

In this section, we consider an application of the formulation in the 
the preceding section to weak lensing field, which is a
projected field of the matter density field along the line of sight. We
then test the performance of our method by comparing 
the model predictions with ray-tracing simulations. Note that the
following formulation can be applied to any projected field such as the
thermal Sunyaev-Zel'dovich effect.

\subsection{From 3d to 2d: lensing convergence and halo sample variance} 

The 
lensing convergence field in angular direction $\vtheta$ on the sky and
for a source galaxy at redshift $z_s$
is given by the weighted projection of the matter density field along
the line of sight:
\begin{equation}
\hat{\kappa}(\vtheta) = \int_0 ^{\chi_S}\!\! d\chi \, q(\chi,\chi_s) 
\hdelta[\chi, d(\chi)\vtheta] ,
\label{eq:kappa}
\end{equation}
where $\chi$ refers to the radial comoving distance, $\chi_s$ is the
distance to the source, and $d(\chi)$ is the comoving angular diameter
distance. 
The function
$q(\chi,\chi_s)$ is the lensing projection kernel defined as
\beq
q(\chi,\chi_s) \equiv  \frac{3}{2} \left( \frac{H_0}{c} \right)^2 \frac{\Omega_m}{a(\chi)} \frac{d(\chi) d(\chi_S - \chi)}{d(\chi_S)},
\eeq
where $\Omega_m$ is the present-day energy density parameter of
matter.

Employing
the Limber approximation, 
we express 
the $n$-point correlation function of the convergence field as the
line-of-sight projection of the corresponding matter correlation function:
\beq
\label{eq:limber}
\bal
\bar{P}^{\kappa}_N (\vl_1, ..., \vl_N) 
&=
 \int_0^{\chi_S}\!\!d\chi \,\,\,
 \frac{q^N(\chi)}{d^{2(N-1)}(\chi)} \,\,\, 
 \bar{P}^\delta_N\!\! \left(  \vk_1 =\frac{\vl_1}{d(\chi)}, ..., \vk_N =\frac{\vl_N}{d(\chi)}; \chi  \right) .
 \eal
\eeq
The lensing power spectrum and bispectrum are obtained for $N=2$ and
$3$, respectively.
Ref.~\cite{LoverdeAfshordi:08} showed that 
the Limber approximation holds a good approximation 
for $l\gtrsim 100$ in which we are most interested.
In the following, we consider a flat-geometry universe for
 simplicity for which we can use the relation 
$d(\chi)=\chi$.

We consider a survey with finite area
$\Omega_S$. 
For the finite-volume effect on a projected density field, we
need to consider super-survey modes at each redshift along the line of
sight. We simply discretize the survey volume into volume elements at
each redshift; $dV(z)\equiv \chi^2\Omega_S\Delta \chi$, where $\chi$
is the comoving distance to redshift $z$ and $\Delta\chi$ is the
width. In this setting, we can define the coherent density mode across
the volume element around redshift $z$, $dV(z)$, as
\beq
\delta_b(z) \equiv 
\int d^3 \vx \,\,\, W(\vx; z) \hdelta(\vx) ,
\eeq
where $W(\vx; z)$ is the window function of the volume element
$dV(z)$. We simply assume a circular-aperture, cylinder-shape geometry
for $dV(z)$;
$W(\chi',\chi\vx_\perp)=W_\parallel(\chi')W_\perp(\vx_\perp)$ with
{$W_\parallel = \Theta(1-2|\chi'-\chi|/\Delta\chi) / \Delta\chi$} and 
{$W_\perp=\Theta(1-x_\perp/\chi\theta_S) / \left(\chi^2\Omega_S\right)$}. Here $W_\parallel(\chi')$ and
$W_\perp(\vx_\perp)$ are the window functions parallel or perpendicular
to the line of sight direction, $\theta_s$ is the survey size
($\Omega_S=\pi \theta_s^2$), and $\Theta(x)$ is the Heaviside step
function, defined such that $\Theta(x)=1$ if $x>0$, otherwise
$\Theta(x)=0$. Then the variance of the 
average
density fluctuation is
given as
\beq
\label{eq:dsigma}
\left\langle {\delta}_b(z)^2\right\rangle
\equiv \frac{1}{\Delta\chi}d\sigma^2(z;\Omega_S)
\simeq \frac{1}{\Delta\chi}
\int \frac{d^2 \vk_\perp}{(2\pi)^2} | W_\perp(\vk_\perp) |^2 
P_{\rm lin}(| \vk_\perp |;\chi) ,
\eeq
where $W_\chi(k_\perp)=2J_1(k_\perp\chi\theta_S)/(k_\perp\chi\theta_S)$
and we have assumed that the radial bin width $\Delta\chi$ is narrow
compared to $\chi$.

\subsection{Ray-tracing simulations and halo model ingredients}
\label{sec:sim}

To test the analytical model we developed in this paper,
we use ray-tracing simulations in Sato et al. \cite{Satoetal:09}. In
brief, the simulations were generated based on the algorithm in
Ref.~\cite{HamanaMellier:01},  
using N-body simulation outputs of large-scale structure 
for a $\Lambda$CDM universe
that is characterized by 
$h=0.732$ ($H_0=73.2$ km s$^{-1}$ Mpc$^{-1}$), $\Omega_m=0.238$,
$\Omega_b=0.042$, and the linear matter power spectrum with 
$n_s=0.958$ and $\sigma_8=0.76$. 
In this paper, we use the simulation results for source redshift $z_s=1$
and use the 1000 realizations of simulated convergence maps and the
friend-of-friend halo catalogs \citep[also see][]{TakadaSpergel:13},
where each realization has an area of $5\times 5=25$ square degrees. We
estimate the power spectrum and bispectrum from 1000 realizations
following the method in Ref.~\cite{Kayoetal:12}. The estimated power spectra
and bispectra were considered to be accurate to within about 5\% in the
amplitudes up to $l\sim 6000$ or $4000$, respectively 
\citep{Satoetal:09,Kayoetal:12}. We also use the 1000
realizations to estimate the covariances and the cross-covariances for
the power spectra, bispectra and halo number counts.

The ray-tracing simulations we use were done in a light cone
volume with an observer's position being its cone vertex. The ray-tracing
simulations include contributions from N-body Fourier modes with length
scales greater than the light-cone volume at each lens redshift (see
Fig.~1 in Ref.~\cite{Satoetal:09}). Thus the simulations are suitable to
study the HSV effect.

As for the halo model, we need to specify its ingredients to
compute the model predictions for the same cosmological model as that of
the simulations.
We employ the Sheth-Tormen fitting formula to compute the halo
 mass function \cite{ShethTormen:99}, for which we employ the parameter
 $q=0.75$ instead of the original value $q=0.707$ according to the
result in Ref.~\cite{HuKravtsov:03}. Similarly, we use the linear halo bias
for the Sheth-Tormen mass
function \cite{Moetal:97,CooraySheth:02}.
We use the formula in Ref.~\cite{Bullocketal:01} to compute 
the linear-theory extrapolated overdensity $\delta_c$ 
for halo formation, and use the formula in Ref.~\cite{Bryanetal:98} for the
virial overdensity $\Delta_\text{vir}$. We employ the Navarro-Frenk-White
model \cite{Navarroetal:97} for the halo profile, where we assume the halo mass and
concentration parameter relation given in Ref.~\cite{TakadaJain:03}.
We have checked that the halo model predictions for the lensing power
spectrum and bispectrum are in reasonably good agreement with the
simulation results to 10--20\% accuracy in their amplitudes over the range
of multipoles we consider.

\subsection{Cluster counts and convergence $n$-point functions} 

\subsubsection{Angular cluster counts} 

We assume that our hypothetical survey gives us access to massive
clusters in the light-cone volume, and that the angular number counts of
clusters can be estimated from the data. The cumulative, angular number
count of clusters 
in the $i$-th mass bin 
up to redshift $z_b$ is given
by an integration of halo mass function over the light-cone volume:
\begin{equation}
 \bar{N}^{2D}_{i,<z_b}\equiv 
\left\langle\hat{N}^{2D}_{i,<z_b}\right\rangle=
\Omega_S\int_0^{\chi_b}\!\!d\chi~ \chi^2
\bar{n}_i(\chi),
\end{equation}
where $\chi_b\equiv \chi(z_b)$,  and we considered a simple survey
geometry (ignored any masking effect for simplicity).
From Eq.~\eqref{eq:number_counts}, the covariance matrix is found
to be
\beq
\bal
\text{Cov} [  \hat{N}_{i, < z_b}, \hat{N}_{j, < z_b'} ]
&= \deltak_{i,j} \hat{N}_{i, <\text{min}(z_b, z_{b'})} 
+\Omega_S^2 \int_0^{\text{min}(\chi_b, \chi_{b'})} d\chi \,\,\, \on_i
\on_j b_i b_j\chi^4 d\sigma^2 (\chi; \Omega_s) .
\eal
\eeq 
Here $d\sigma^2(z;\Omega_S)$ is given by Eq.~(\ref{eq:dsigma}). The
above expression matches the result in Ref.~\cite{TakadaSpergel:13}.

\subsubsection{Covariance for lensing 
$n$-point correlation functions}

Similarly to the discussion in Section~\ref{sec:n-point}, the HSV
contribution to the covariance between the $n$- and $n'$-point
correlation functions of the convergence field,
$\hat{P}^\kappa_n$ and $\hat{P}^\kappa_{n'}$,
 is given under our formulation as
\beq
\label{eq:hsvpn_2d}
{\rm Cov}[\hat{P}^\kappa_n,\hat{P}^\kappa_{n'}]^{\rm HSV}
=
\int_0^{\chi_S}\!\! d\chi \,\,\, \frac{q^{n+n'}(\chi) }{\chi^{2(n+n'-2)}}
\frac{\partial \bar{P}_n}{\partial \delta_b}
\frac{\partial \bar{P}_{n'}}{\partial \delta_b}
d\sigma^2 (\chi; \Omega_S) .
\eeq
Note that we can include the effect of the coherent super-survey
mode $\delta_b$ on the 1-halo term and the different halo terms by
computing the response functions such as $\partial P_n/\partial
\delta_b$. This differs from what was done in the previous study, where 
only the 1-halo contribution was computed.

\begin{figure}
\begin{center}
\includegraphics[width=3.5in]{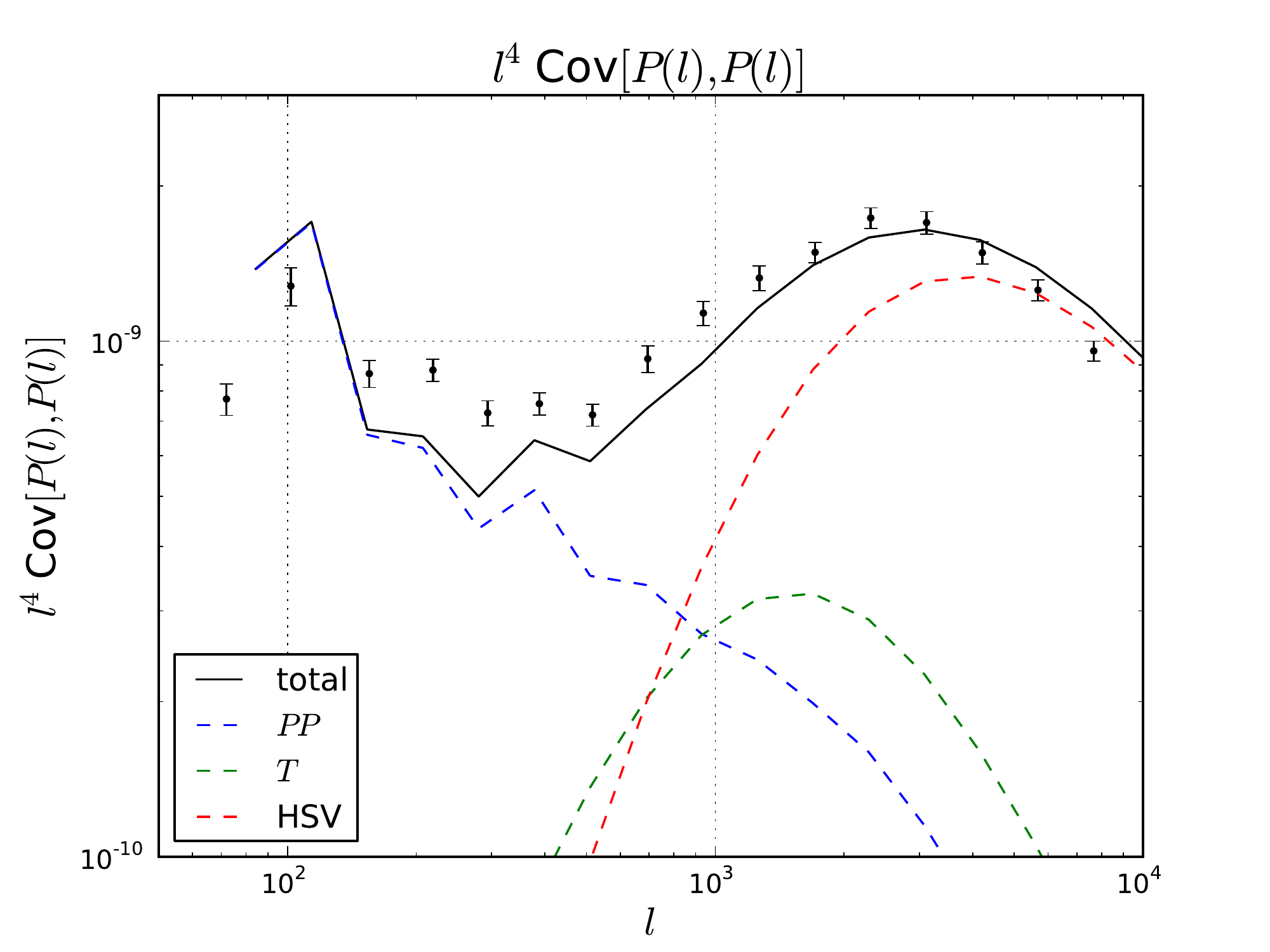}
\includegraphics[width=3.5in]{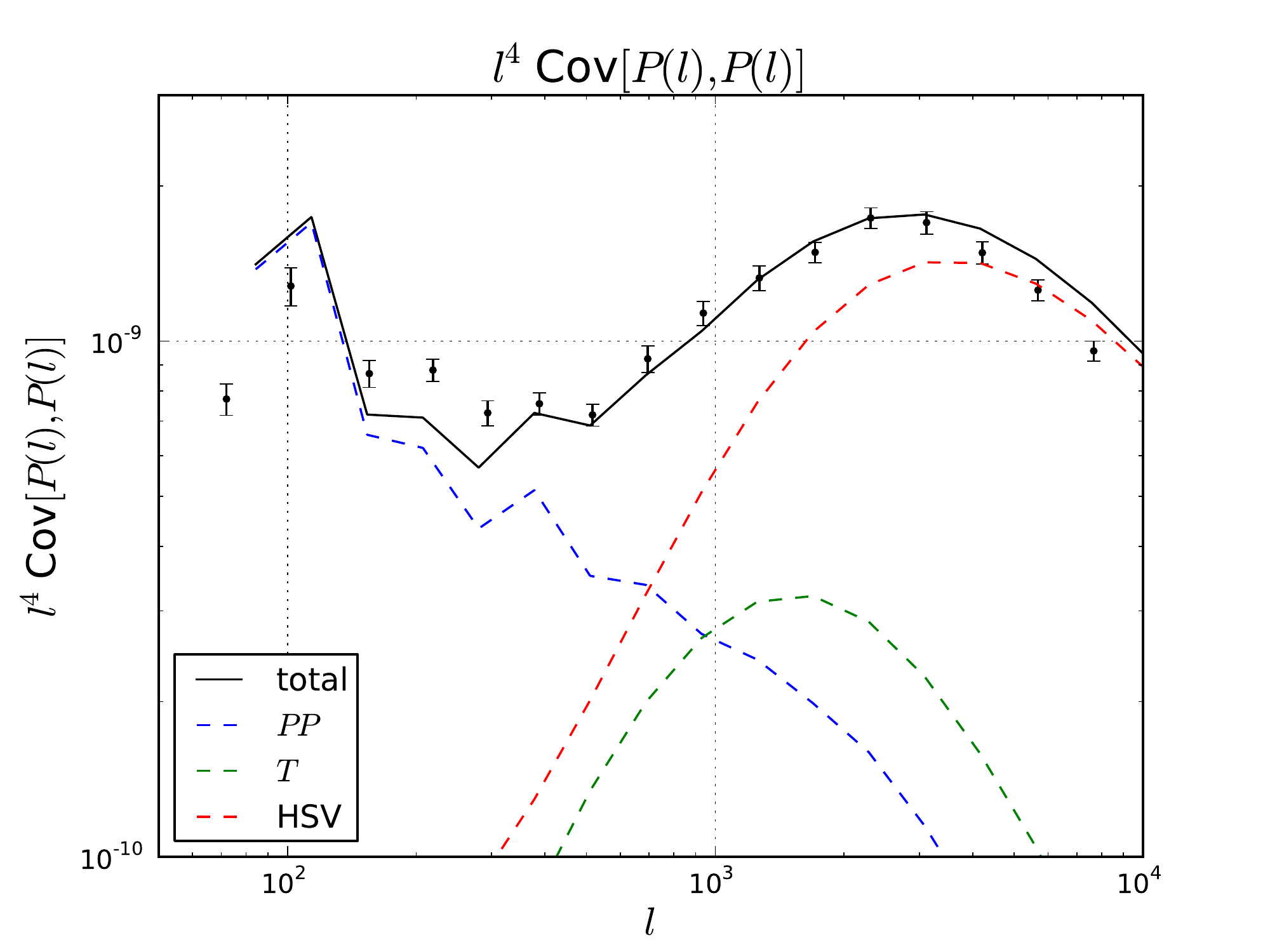}
\caption{Comparison of our analytical prediction for the lensing power
spectrum covariance (Eqs.~\ref{eq:hsvpn_2d} and \ref{eq:covp}) 
with the ray-tracing simulation results obtained from 1000
 realizations. Here we considered a single source redshift $z_s=1$.
For the analytical predictions,  
the Gaussian term
(PP), the trispectrum term (T) and the halo sample variance (HSV) are
included. 
The error bars for the simulation results are obtained from the variance
 of the 1000 realizations, which correspond to $1\sigma$ scatters of
 power spectrum estimation
 for 25 square degrees, the area of each ray-tracing simulation. 
\textit{Left panel:} We included only the 1-halo term for the HSV
 calculation. 
\textit{Right:} We further included the 2-halo term contribution for the
 HSV effect. 
The analytical predictions are in fairly good agreement with the
 simulation results, over a wide range of multipoles, {\em if} the HSV
 effect is included in the analytical prediction. Comparing the left and
 right panels shows that including the 2-halo term of the HSV effect
improves the agreement at the transition regime between the 1- and
 2-halo terms, in the range of $l\simeq $a few hundreds to $10^3$.
}
\label{fig:covp}
\end{center}
\end{figure}
\begin{figure}
\includegraphics[width=3.6in]{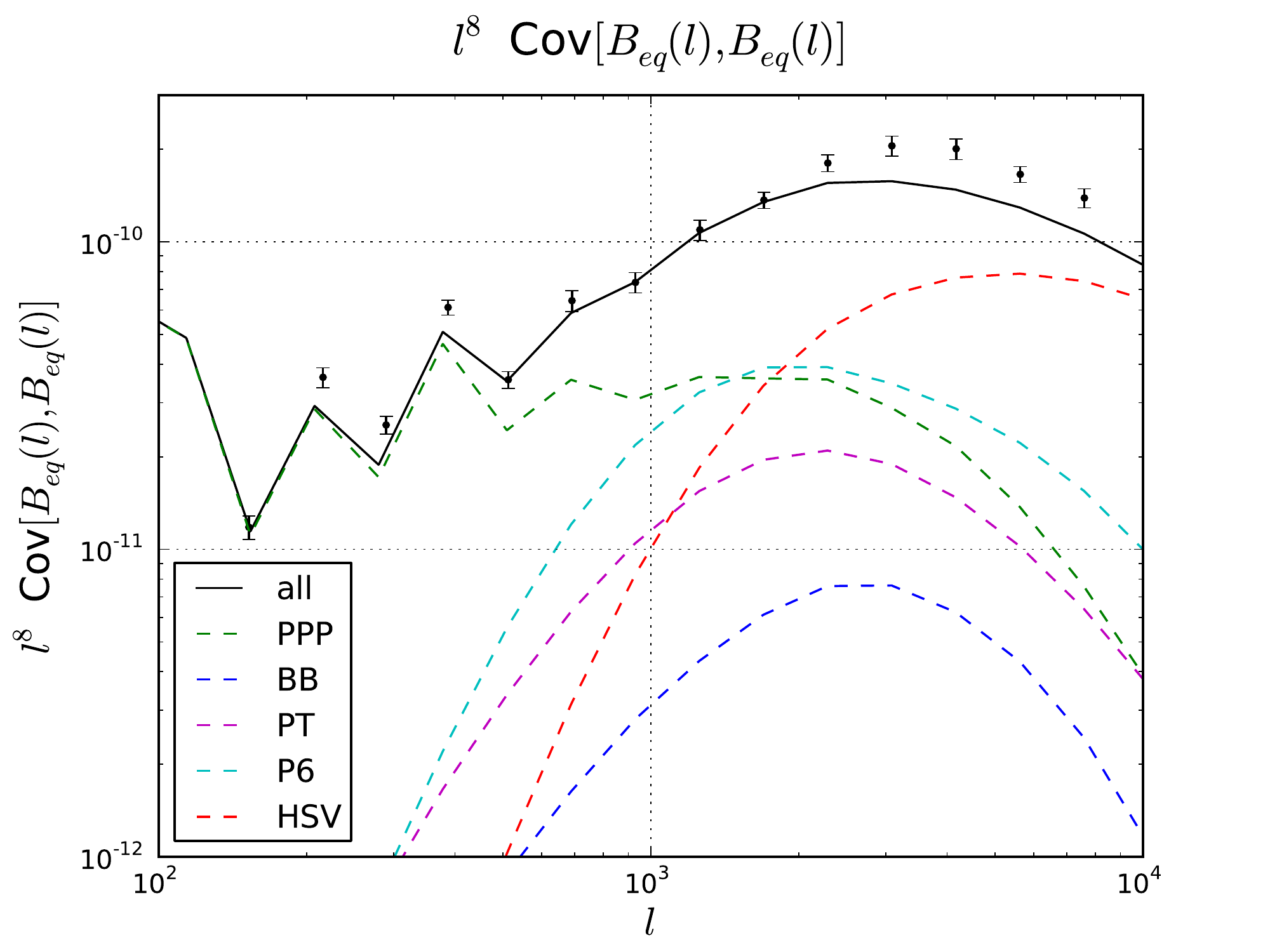}
\caption{Comparison of the halo model prediction with the simulation
 results for the lensing bispectrum covariance for equilateral triangle
 configurations, as a function of the side length.  For the analytical
 model, we included
the standard Gaussian and non-Gaussian contributions arising from a
combination of the correlation functions up to the 6-point correlation
function (PPP, BB, PT, P6, from Eq~\ref{eq:std_covb}) and also included the HSV contribution
 (Eq.~\ref{eq:hsvpn_2d}).
 For the HSV contribution, we included all 1-, 2- and 3-halo terms, 
 but only the 1=halo term gives an important contribution.
} \label{fig:covb}
\end{figure}
\begin{figure}
\includegraphics[width=3.6in]{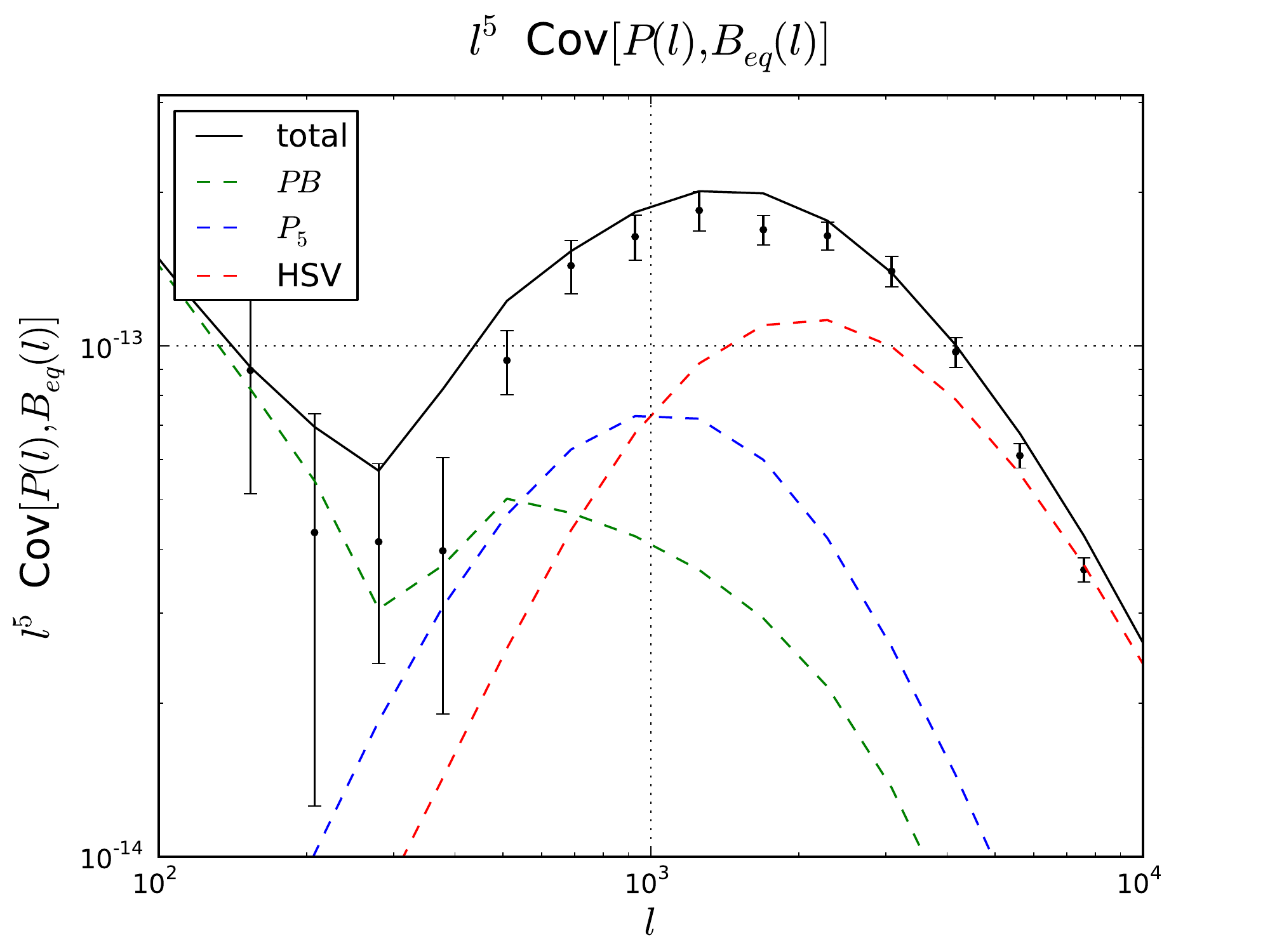}
\caption{Cross-covariance between the lensing power spectrum,
 $P_\kappa(l)$,  
and the 
 bispectrum of equilateral triangle configuration, $B_{\rm eq}(l,l,l)$,
 against multipoles. Similarly to the previous plot, for the halo model
 prediction, we included the standard contributions
(PB, P5, from Eq.~\ref{eq:std_covpb}) and the HSV contribution 
 (Eq.~\ref{eq:hsvpn_2d}). 
}
\label{fig:covpb}
\end{figure}
Fig.~\ref{fig:covp} shows the diagonal elements of the power spectrum
covariance as a function of multipole. The halo model predictions are in
fairly good agreement with the simulation results. This agreement can
be realized only if including the HSV contribution. The right panel
shows the results for the halo model when including the HSV
contributions for the 2-halo term, which can be compared with the
previous study such as Ref.~\cite{Satoetal:09}. The figure shows that
including the HSV 2-halo term improves the agreement over a range of the
transition regime between the 1- and 2-halo terms.  Note that these
results are for survey area of 25 sq. degrees, the area of the
ray-tracing simulations we use (see Section~\ref{sec:sim}), but the HSV
effects are significant for any survey area of upcoming surveys (see
Refs.~\cite{Kayoetal:12,TakadaHu:13}).

In Figs.~\ref{fig:covb} and \ref{fig:covpb}, we show the results
for the bispectrum covariance and the cross-covariance between power
spectrum and bispectrum. We followed the method in
Ref.~\cite{Kayoetal:12}, and for both the figures we considered the
bispectra of equilateral triangle configurations against the side
length.  The halo model is again in good agreement with the simulations,
to a level of 10--20\% accuracy in their amplitudes.

\subsubsection{Cross-covariances between angular number counts of halos
   and the 
lensing $n$-point correlation functions} 

Applying the formulation for 3D fields in
Section~\ref{sec:ccov3D} to 2D fields, we can estimate the
cross-covariance between the angular number counts of clusters 
and the
lensing power spectrum:
\begin{eqnarray}
\label{eq:covnp_2d}
\text{Cov} [  \hat{N}_{M>M_{\rm th}}, \hat{P}^\kappa(l)  ]
&=& 
\int_0^{\chi_S} d\chi \,\,\, \frac{q^{2} (\chi) }{\chi^{2}}
\left[
\sum_{i>i_{min}} \bar{n}_i p_i^{1h}(k)
+
2 \sum_{i>i_{min},j} \bar{n}_i \bar{n}_j  p_{ij}^{2h}(k)
\right] 
\nonumber \\
&&
+ \Omega_S 
\int_0^{\chi_S} d\chi \,\,\, 
q^{2} (\chi)
\left( \sum_{i>i_{min}} 
\on_i b_i \right)
\frac{\partial \bar{P}(k)}{\partial \delta_b}
d\sigma^2 (\chi;\Omega_S),
\end{eqnarray}
where $k=l/\chi$ in the arguments on the r.h.s.

The cross-covariance for the lensing bispectrum is
\begin{eqnarray}
\label{eq:covnb_2d}
\text{Cov} [  \hat{N}_{M>M_{\rm th}}, \hat{B}(\vec{l})  ]
&=&
 \int_0^{\chi_S} d\chi \,\,\, \frac{q^{3} (\chi) }{\chi^{4}}
\left[
\sum_{i>i_{min}} \bar{n}_i b_i^{1h}(\vec{k})
+
2 \sum_{i>i_{min},j} \bar{n}_i \bar{n}_j  b_{ij}^{2h}(\vec{k})
+
3 \sum_{i>i_{min},j,l} \bar{n}_i \bar{n}_j \bar{n}_l  b_{ijl}^{3h}(\vec{k})
\right] \nonumber \\
&&+ \Omega_S 
\int_0^{\chi_S} d\chi \,\,\, 
\frac{q^{3} (\chi) }{\chi^{2}}
\left(\sum_{i>i_{min}} 
\on_i b_i \right)
\frac{\partial \bar{B}(\vec{k})}{\partial \delta_b}
d\sigma^2 (\chi;\Omega_S), 
\end{eqnarray}
where we have again used the collapsed notation such as
$\vec{l}=(l_1,l_2,l_3)$ and $\vec{k}=(l_1/\chi,l_2/\chi,l_3/\chi)$.

To be more general, the cross-covariance for the lensing $N$-point
correlation function is given as
\begin{eqnarray}
\text{Cov} [  \hat{N}_{M>M_{\rm th}}, \hat{P}_N  ]
&=& \int_0^{\chi_S} d\chi \,\,\, \frac{q^{N} (\chi) }{\chi^{2(N-1)}}
\left[
\sum_{i>i_{min}} \bar{n}_i p_i^{1h}
+
2 \sum_{i>i_{min},j} \bar{n}_i \bar{n}_j  p_{ij}^{2h}
+... \right.\\
&&\left. +
N
\sum_{i_1>i_{min},i_2,..., i_N} \bar{n}_{i_1} \bar{n}_{i_2}... \bar{n}_{i_N}  p_{i_1, i_2, ..., i_N}^{Nh}
\right] 
+ \Omega_S 
\int_0^{\chi_S} d\chi \,\,\, 
\frac{q^{N} (\chi) }{\chi^{2(N-2)}}
\left( \sum_{i>i_{min}} \on_i b_i \right)
\frac{\partial \bar{P}_N}{\partial \delta_b}
d\sigma^2 (\chi;\Omega_S) . \nonumber
\end{eqnarray}

\begin{figure} 
\includegraphics[width=3.5in]{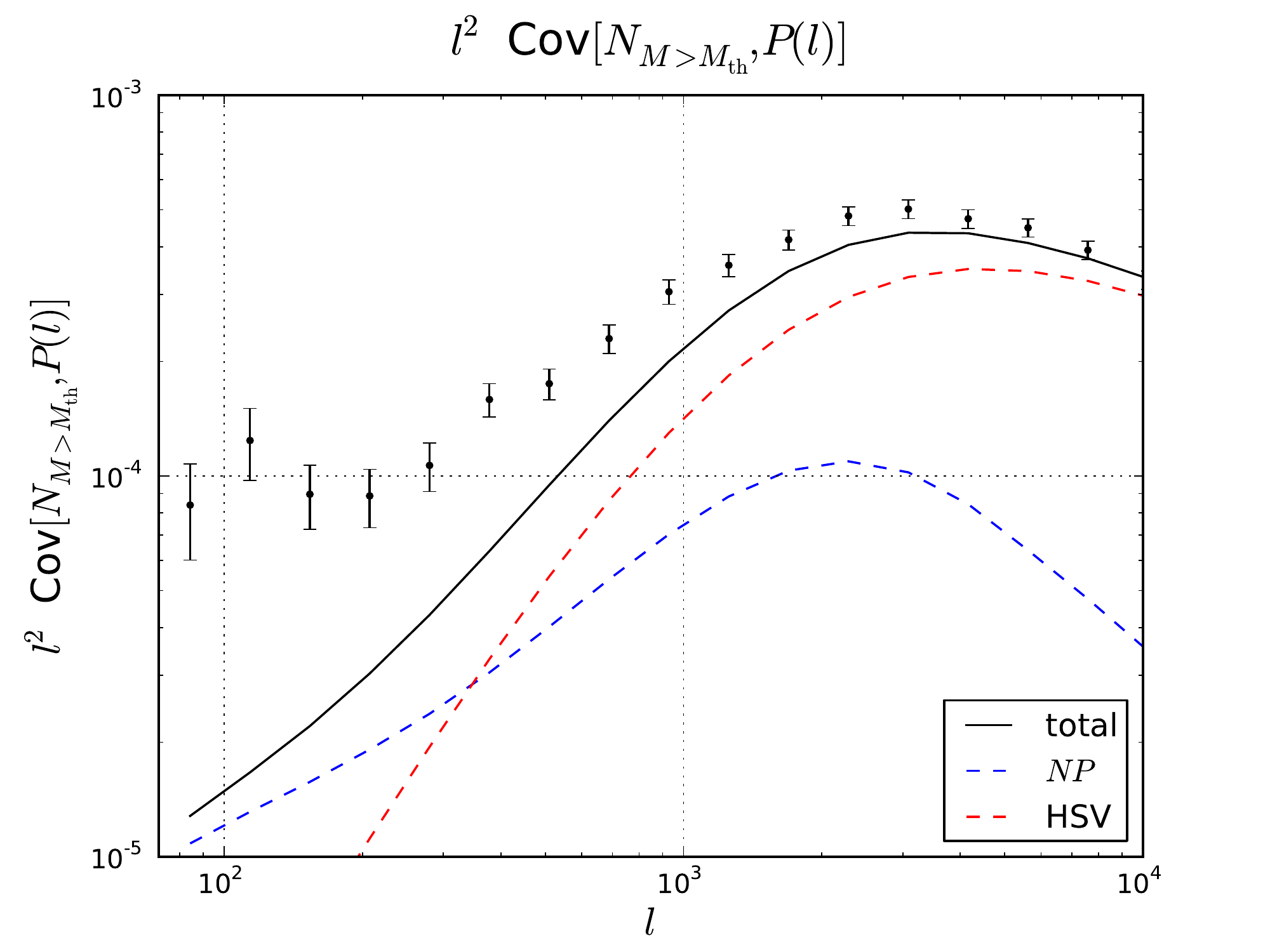}
\includegraphics[width=3.5in]{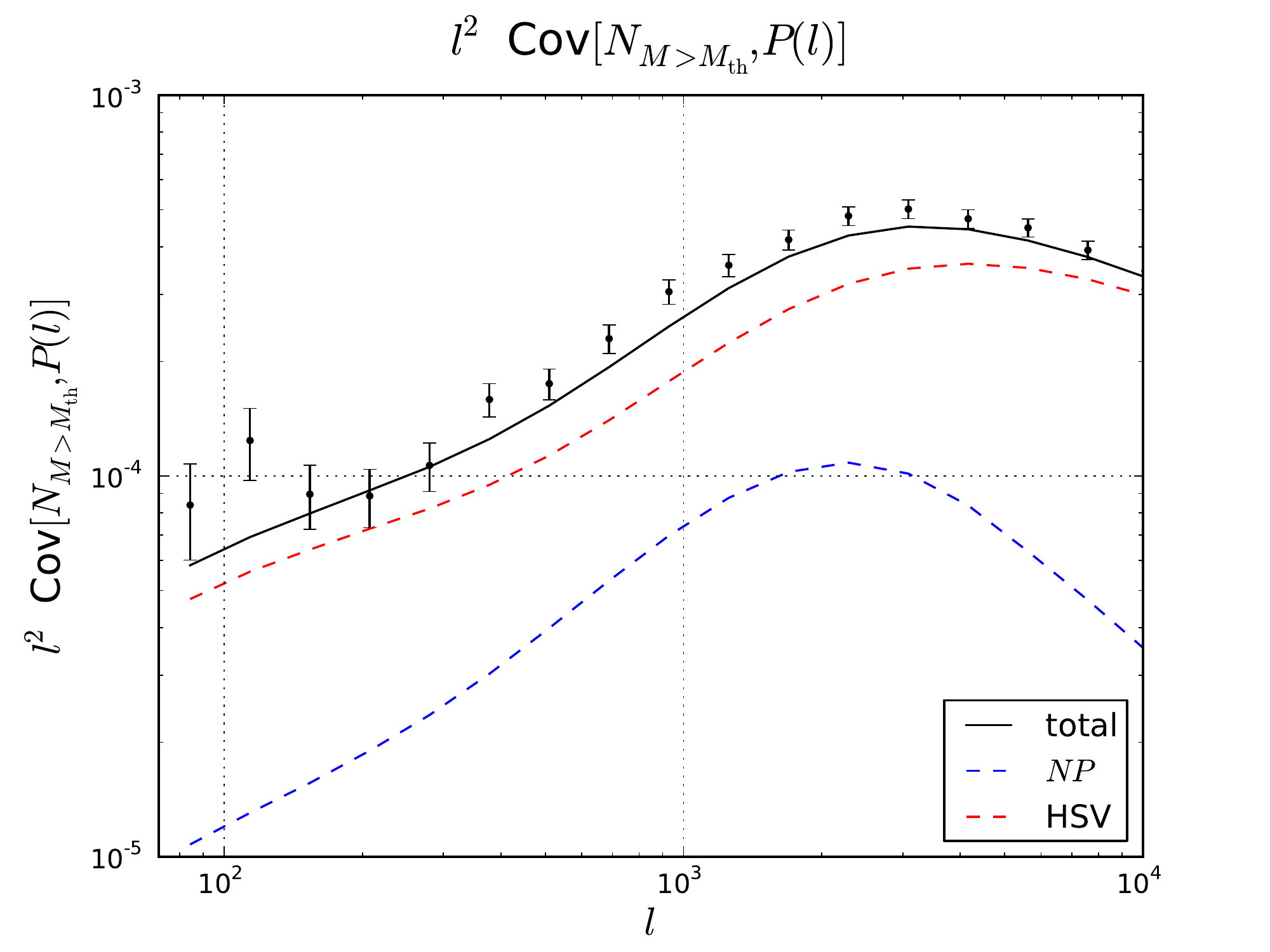}
\caption{Cross-covariance between the angular number counts of halos and
the lensing power spectrum $P_\kappa(l)$ as a function of
multipoles. For the number counts, we included all the halos that are
in the light cone up to $z_s=1$ over area 25 sq. degrees (area of the
ray-tracing simulation) and have masses greater than
$M=10^{14}M_\odot/h$. The error bars for the simulation results are the
variance estimated from the 1000 realizations. For the halo model
prediction, we used Eq.~(\ref{eq:covnp_2d}) to compute the contributions
 arising from the product of the number counts and the power spectrum
 (NP) and the HSV effect. The difference between the left and right
 panels is whether the halo model prediction includes the 2-halo term of
 the HSV effect (right) or not (left). The halo model prediction, with
 the 2-halo HSV effect, well reproduces the simulation result
 over the range of multipoles.
}
\label{fig:covnp}
\end{figure}
\begin{figure} 
\includegraphics[width=3.5in]{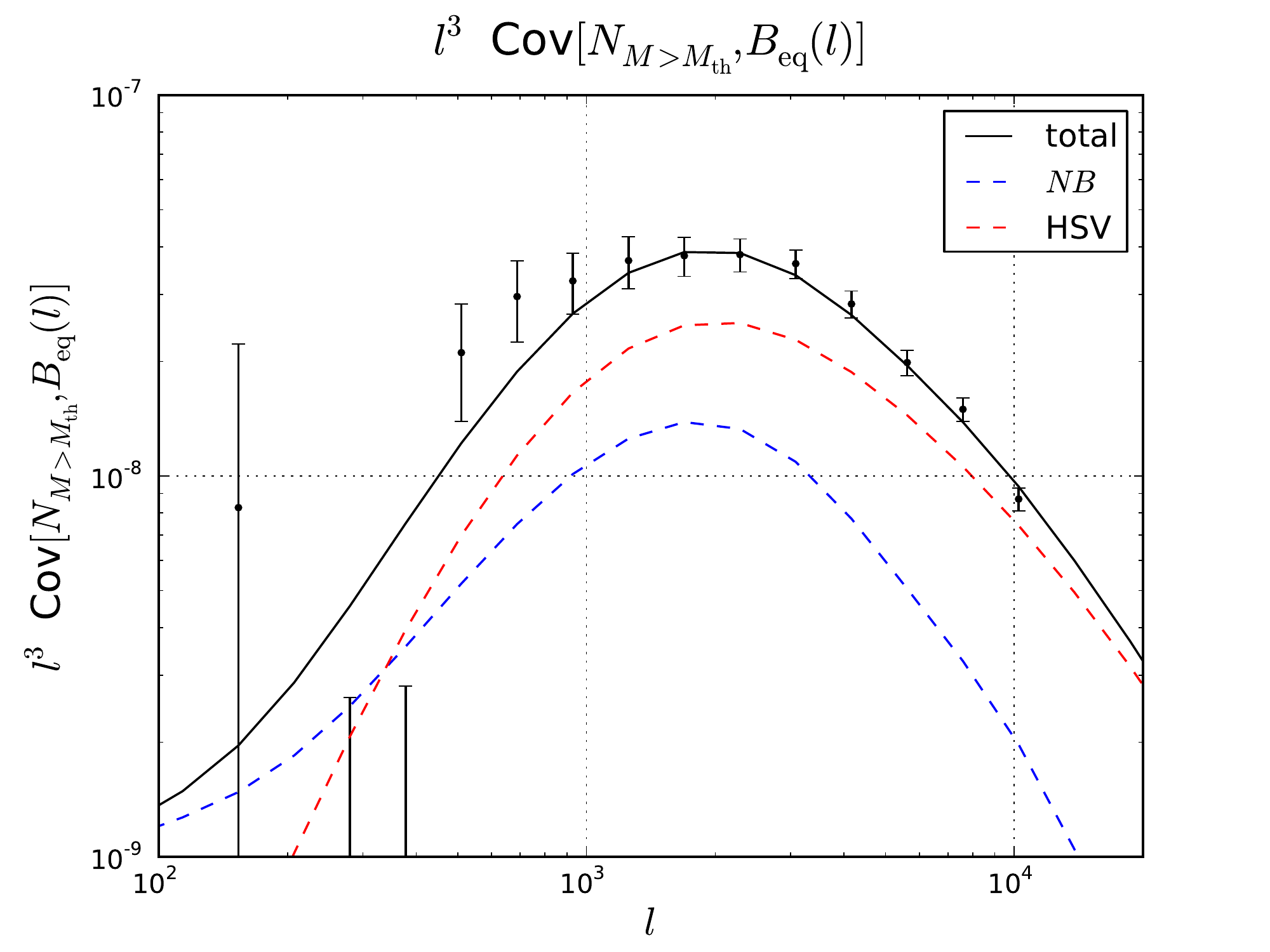}
\includegraphics[width=3.5in]{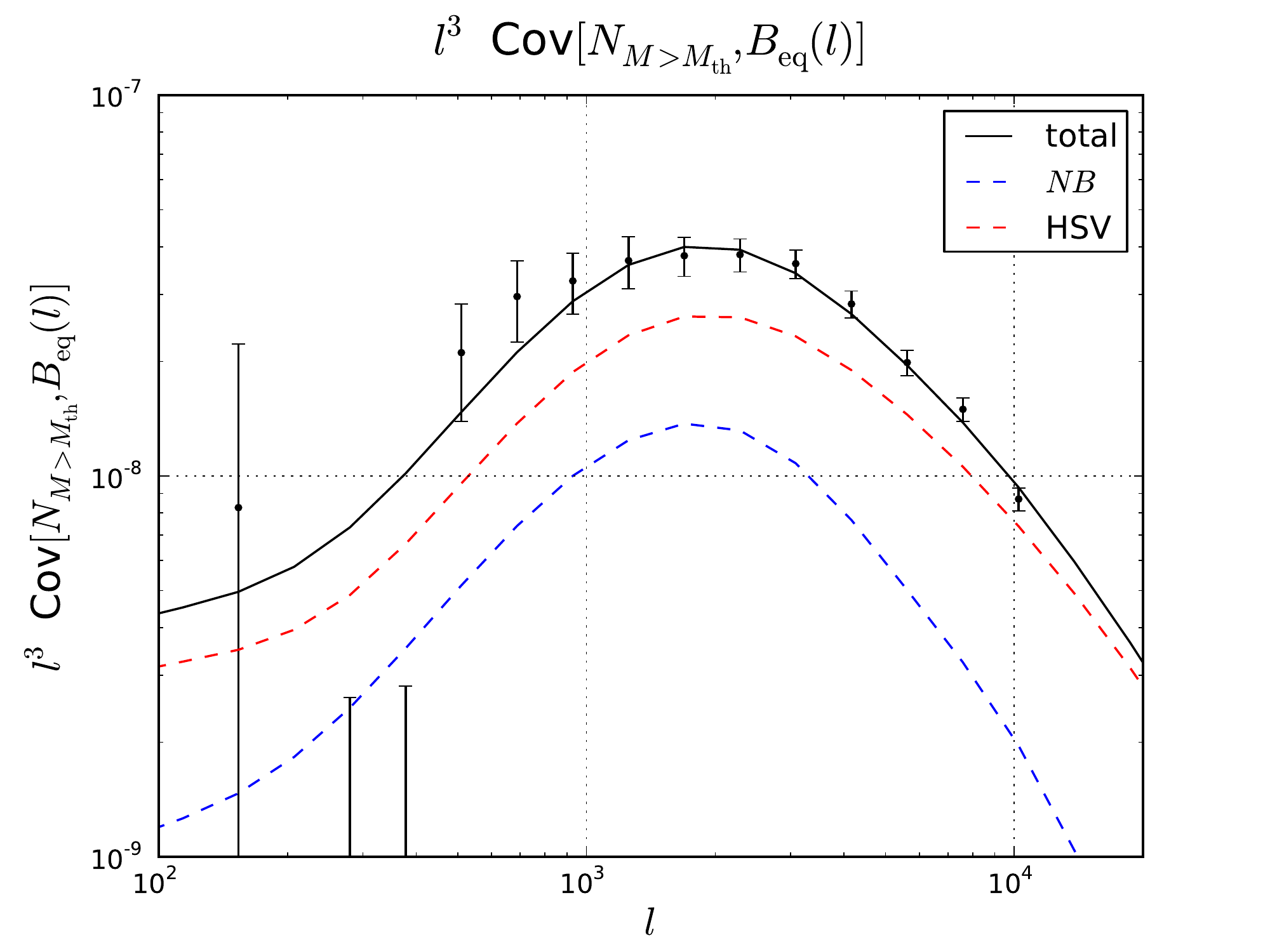}
\caption{
Similarly to the previous plot, the cross-covariance between the
angular halo number counts and the bispectrum of equilateral triangle
configurations, $B_{\kappa, {\rm eq}}(l,l,l)$. The halo model
 computation is based on Eq.~(\ref{eq:covnb_2d}). The difference between
 the left and right panels is whether we included the 2- and 3-halo
 terms for the HSV contribution in the halo model prediction (right
 panel) or not (left). 
}
\label{fig:covnb}
\end{figure}
 
In Figs.~\ref{fig:covnp} and \ref{fig:covnb}, we compare the halo
model predictions for the cross-covariance of the angular number counts
of halos with the lensing power spectrum or bispectrum. For the halo
number counts, we included all the halos that are in the light cone up
to $z_s=1$ and over area of 25 square degrees (area of the ray-tracing
simulation). Both figures show that the halo model predictions are
in fairly nice agreement with the simulation results, if including the
HSV contribution. It is also shown that including the different halo
terms of the HSV effect better agrees with the simulation results over a
range of the transition regime of multipoles between the 1-halo term
and the different halo terms.

\section{Joint likelihood for power spectrum, bispectrum and cluster counts, and Fisher forecast}
\label{sec:likelihood_fisher}

We have so far derived the co- or cross-variances between the halo
number counts, the power spectrum and the bispectrum. In this section,
we discuss their joint likelihood function. Exactly speaking, a
derivation of the likelihood function requires a knowledge on all the
higher-order cumulants of the observables beyond the second-order
moments such as the skewness and kurtosis. Here, we instead assume that
the joint likelihood function obeys a multivariate Gaussian function
that is given by the mean values and the second-order variances (co- or
cross-covariances) of the observables, which we have already derived up
to the previous section. The multivariate Gaussian likelihood is
somewhat expected for the lensing fields at high multipoles due to the
central limit theorem, because the lensing field is from a projection of
independent large-scale structure at different redshifts along the
line-of-sight and also because the power spectrum and bispectrum of high
multipoles are from the averages over a large number of Fourier
modes.

Thus we assume that the joint likelihood function for observables
$\mbf{D}$ obeys the following multivariate Gaussian:
\beq
\label{eq:likelihood}
\mathcal{L} \left( \mbf{D} \right) 
\propto 
\text{exp}\left[ -\frac{1}{2}  \left(\mbf{D} - \bar{\mbf{D}}\right)^t    
\mbf{\Sigma}^{-1}    \left(\mbf{D} - \bar{\mbf{D}}\right) \right],
\eeq 
where $\mbf{D}$ denotes the observable vector, e.g. defined a
$\mbf{D}\equiv \left( \{ \hat{P}(l)\}, \{ \hat{B}(l)\}, \hat{N}_{M >
M_{\rm th}} \right)$,
$\bar{\mbf{D}}$ denotes its mean vector, $\mbf{\Sigma}$ its co- or
cross-variance matrix, and $\mbf{\Sigma}^{-1}$ is the inverse matrix.
The vector and matrix notations are intended to mean the summation over
the cluster mass bin (a single bin though here), the multipole bins or
the triangle configurations.

\begin{figure}
\centering
\includegraphics[width=3.6in]{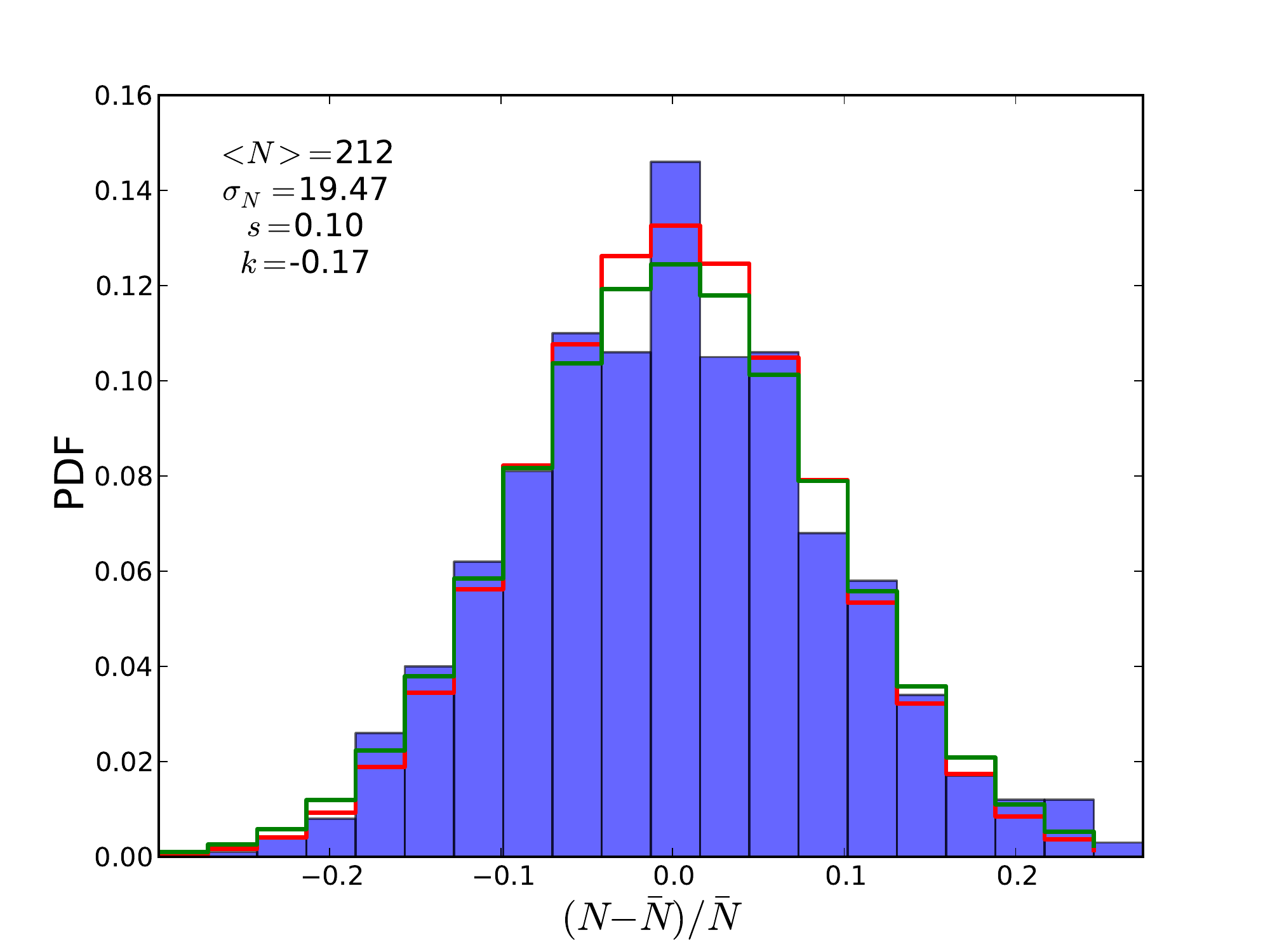}
\caption{
The likelihood distribution of the angular number counts of halos with
 $M\ge 10^{14}M_{\odot}/h$, in the light-cone up to $z_s=1$ and with
 area $25$ sq. degrees. 
The histogram shows
 the distribution measured from the 1000 ray-tracing simulations. 
The red-color, solid curve shows the halo model prediction computed
 assuming the Gaussian likelihood function (Eq.~\ref{eq:likelihood}),
 where we used the halo model to compute the mean and the variance.  
For comparison, the green-color, solid curve shows the Gaussian
 distribution that has the same mean value and variance as those of the
 simulations. The mean, variance, skewness and excess kurtosis values measured from the
 simulations are also given.
}
\label{fig:pdf_n}
\end{figure}
\begin{figure}
\centering
\includegraphics[width=7in]{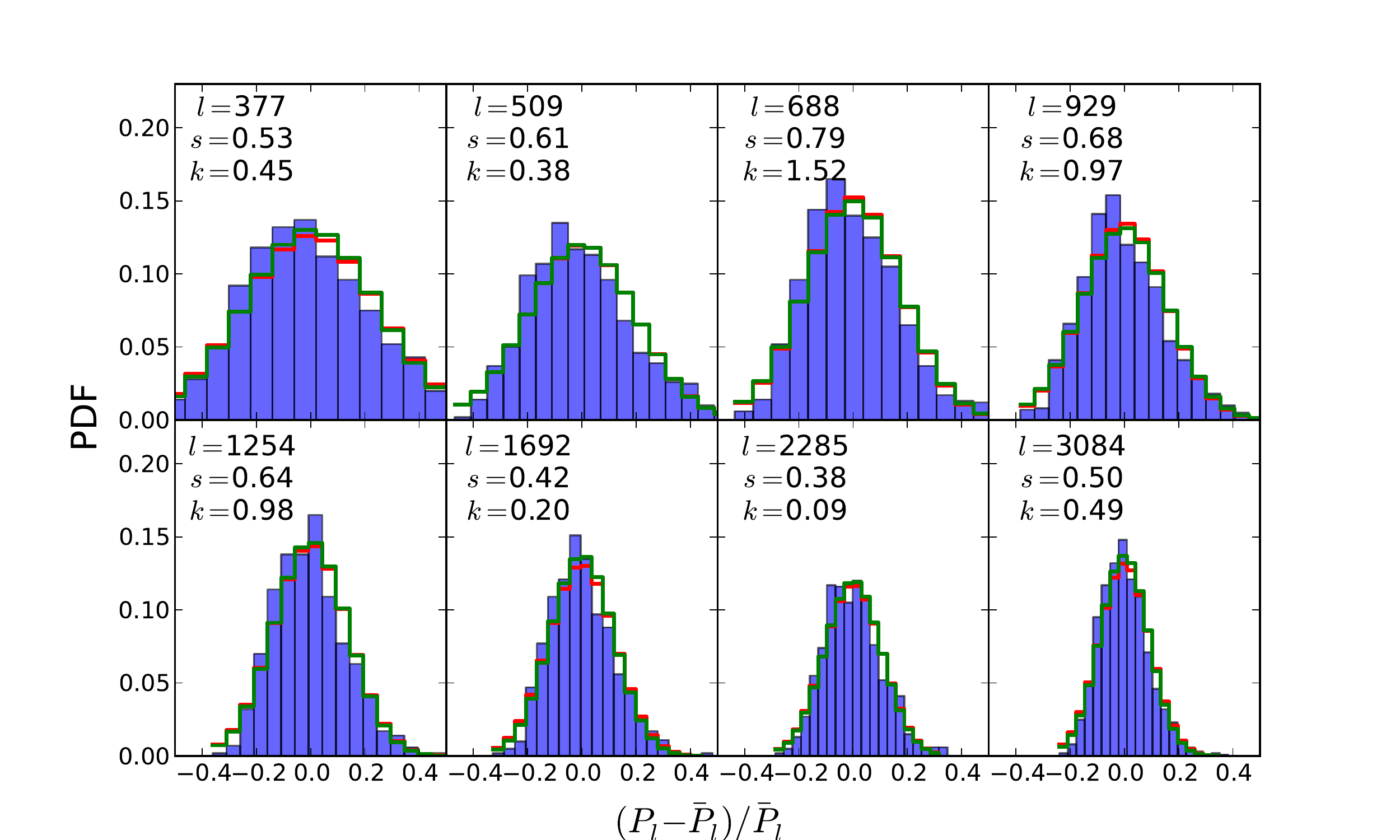}
\caption{Similarly to the previous figure, but for the lensing power
 spectrum $P_\kappa(l)$. 
The different panels show the distributions for different
 multipole bins as indicated. Again the halo model prediction (red, solid
 curve) well reproduces the width of the simulation distribution, if the
 HSV contribution is included. The skewness and kurtosis measured from
 simulations are small compared to the width of the distribution.
}
\label{fig:pdf_p}
\end{figure}
\begin{figure}
\centering
\includegraphics[width=7in]{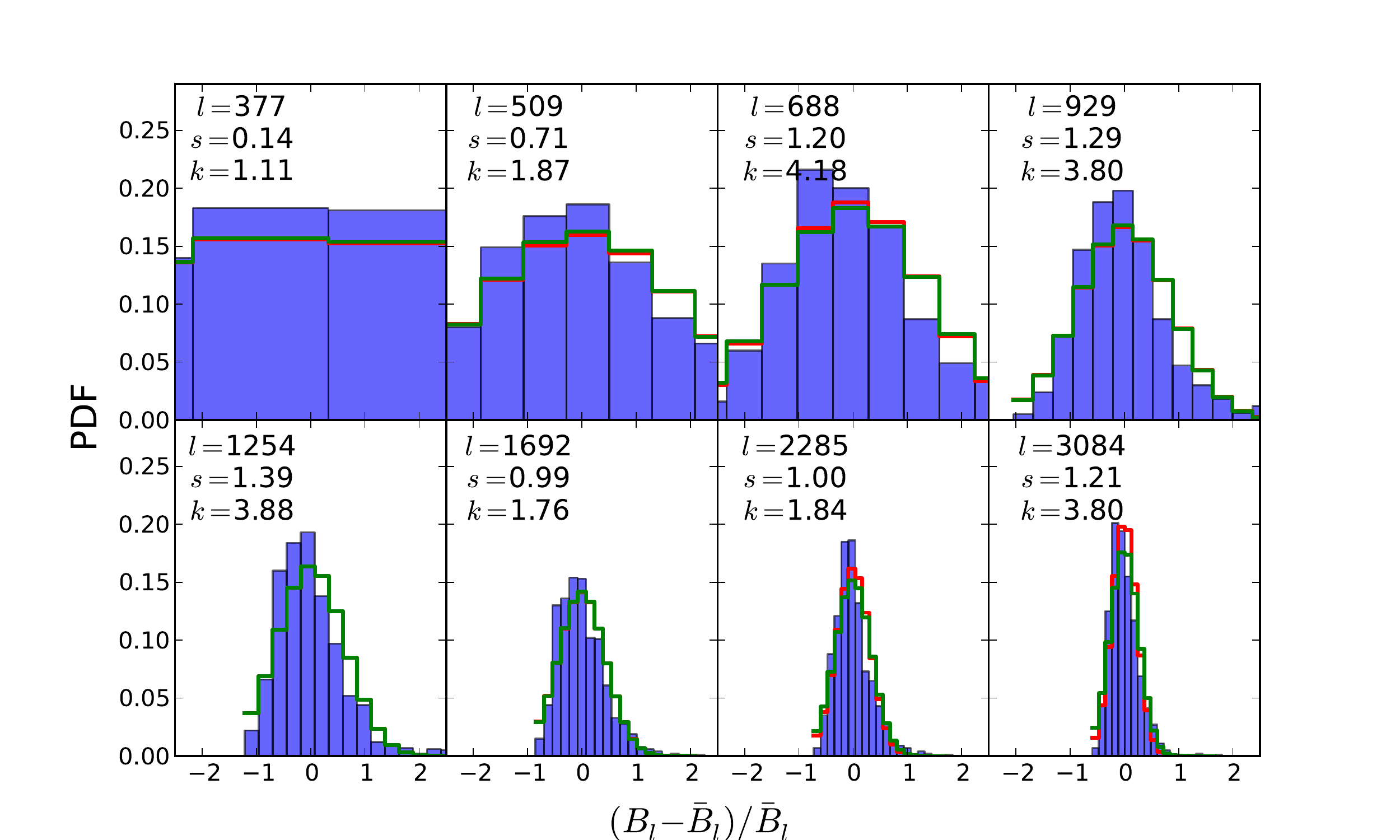}
\caption{
Similarly to the previous plot, but for the lensing bispectrum of
 equilateral triangle configurations. The different panels are for
 different-size triangle configurations. Compared to
 Figs.~\ref{fig:pdf_n} and \ref{fig:pdf_p}, the simulation distribution
 shows a larger asymmetry and therefore larger skewness and kurtosis
 values. The halo model nevertheless well reproduces the width of the
 distribution. 
}
\label{fig:pdf_b}
\end{figure}
In Figs.~\ref{fig:pdf_n}, \ref{fig:pdf_p} and \ref{fig:pdf_b}, we
show the distributions of the angular number counts of halos, the
lensing power spectrum, and the bispectrum of equilateral triangles,
which we measured from the 1000 ray-tracing simulations. Again note that
the distributions are for the area of 25 sq. degrees. These observables
show a fairly symmetric distribution, although the bispectrum shows a
larger skewness than the other two quantities. The red-color solid curve
in each figure shows the halo model prediction
(Eq.~\ref{eq:likelihood}). The halo model appears to well reproduce the
width of the distribution seen in the simulations. The agreement is
realized only if we include the HSV contributions
 as shown in Figs.~\ref{fig:covp} and \ref{fig:covb}.
These figures also
show that the skewness of the distribution is well within the width of
the distribution.

\begin{figure}
\centering \includegraphics[width=7in]{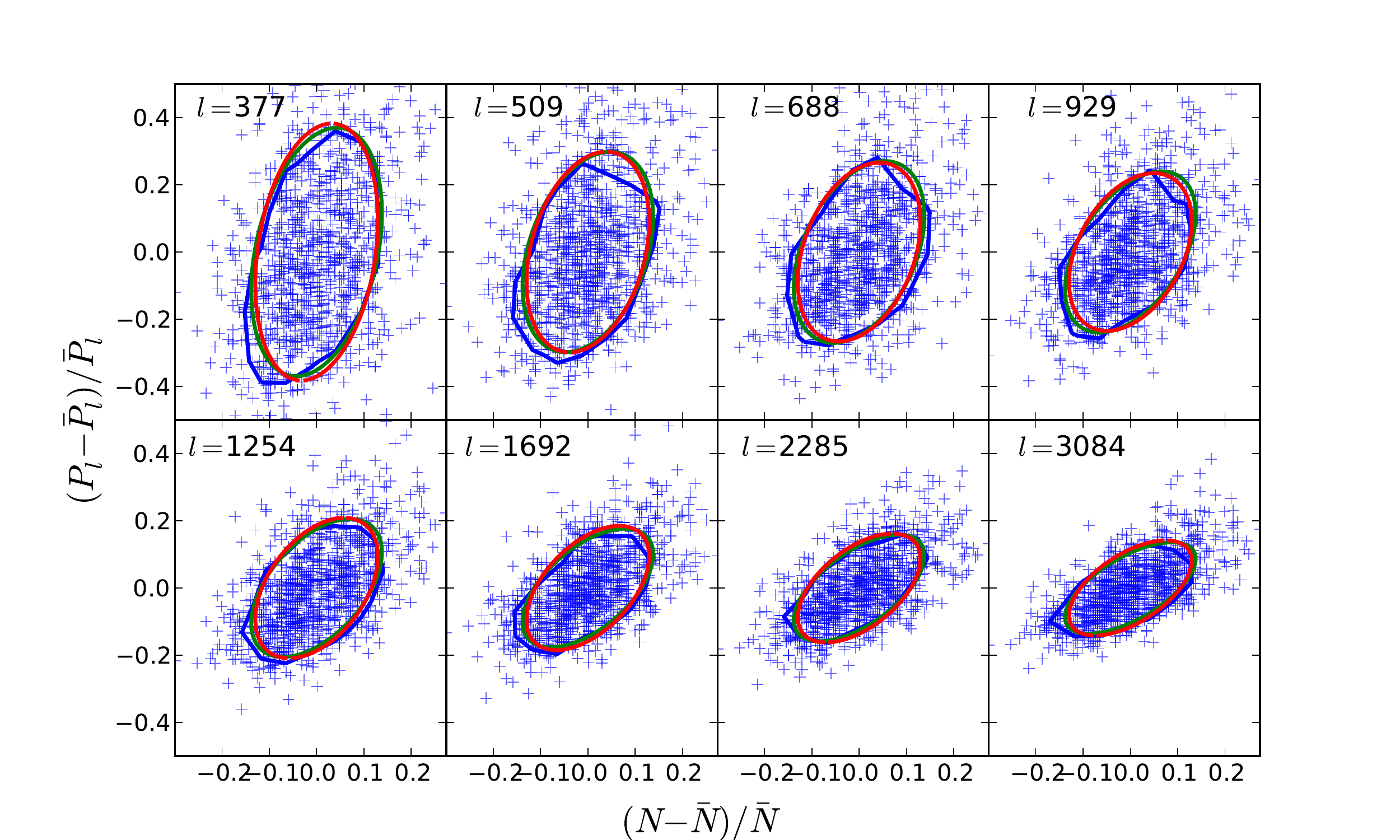}
\caption{A joint
distribution of the angular number counts of halos with $M\ge
 10^{14}M_\odot$ and the lensing power spectrum $P_\kappa(l)$. The
 different panels show the results for different multipole bins
 as denoted.
Each cross symbol denotes one realization out of the 1000 realizations. 
The blue-color contour in each panel shows the 68\% percentile of the
 distribution, which is 
estimated
by binning the 1000 simulations into
 a 2d-histogram.
The red-color contour shows the halo model prediction for
 the joint distribution, computed based on Eq.~(\ref{eq:likelihood}).
For comparison, the green contour shows the multivariate Gaussian
 distribution that has the same mean and variances as those of the
 simulation distribution.
} \label{fig:scatter_np}
\end{figure}
\begin{figure}
\centering
\includegraphics[width=7in]{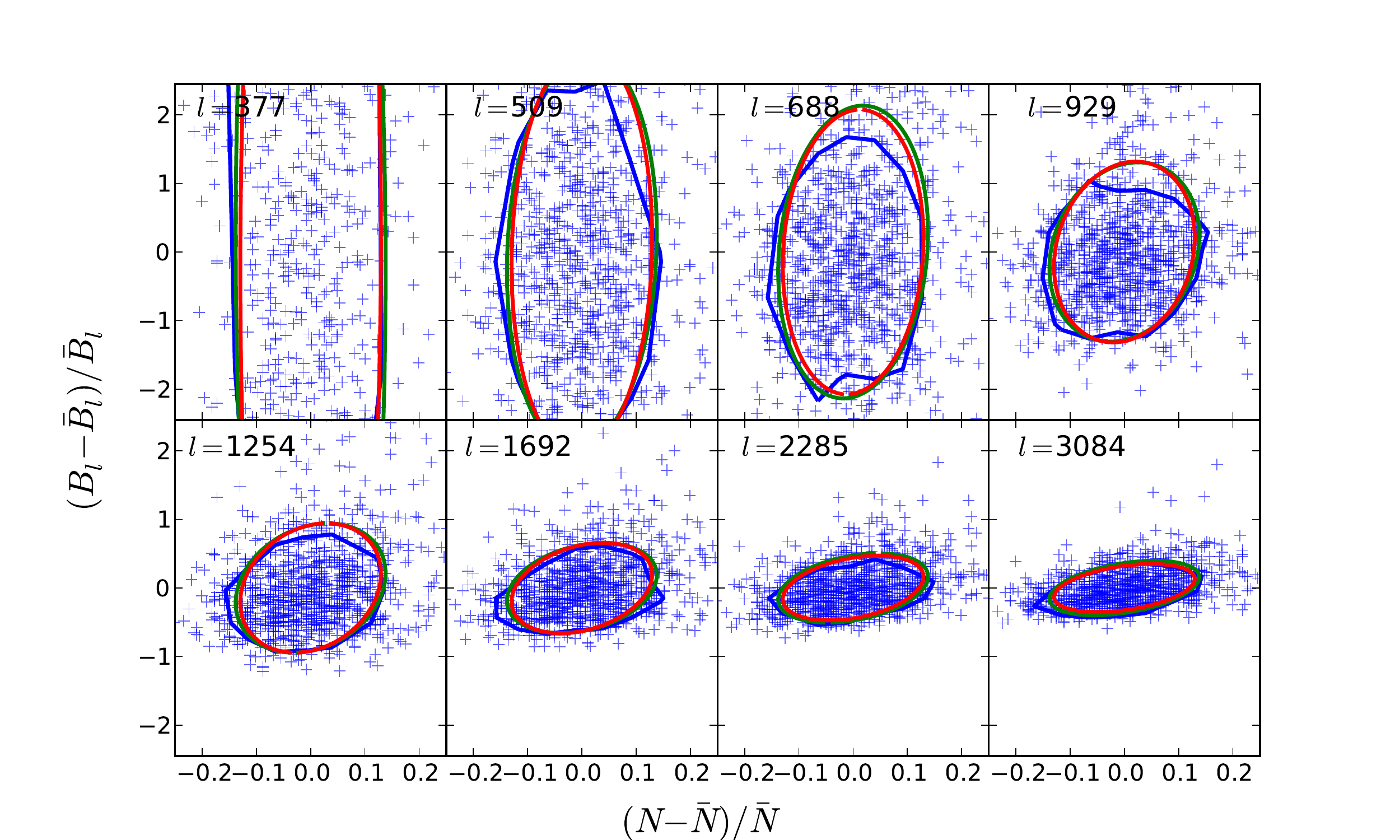}
\caption{
Similarly to the previous figure, but the joint distribution for the
 halo number counts and the lensing bispectrum of equilateral triangle
 configurations, $B_{\kappa, {\rm eq}}(l,l,l)$. The halo model well
 reproduces the distribution seen in the simulations.
}
\label{fig:scatter_nb}
\end{figure}
\begin{figure}
\centering
\includegraphics[width=7in]{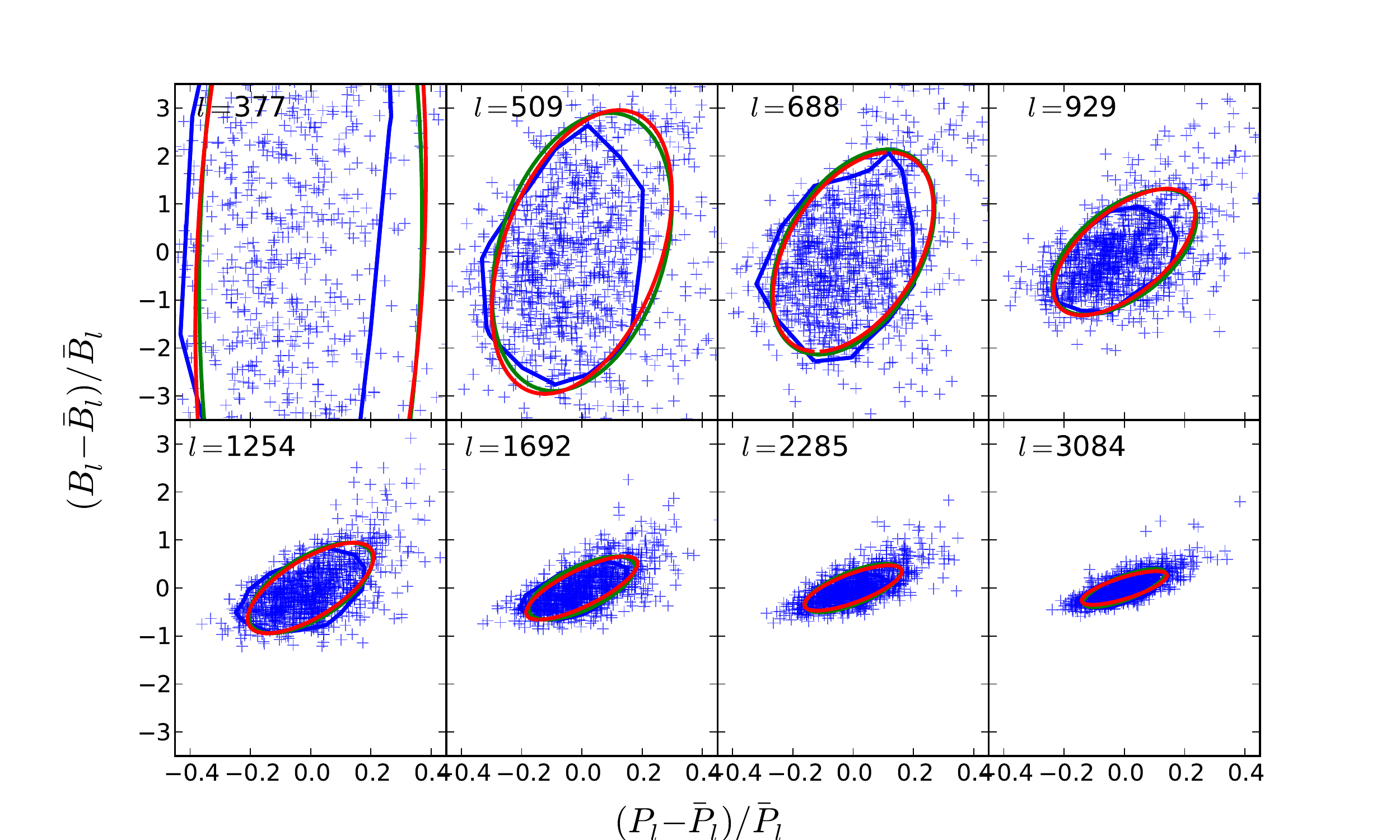}
\caption{Similarly to the previous figure, but the joint distribution
 for the lensing power spectrum and bispectrum. 
}
\label{fig:scatter_pb}
\end{figure}
Figs.~\ref{fig:scatter_np},
\ref{fig:scatter_nb} and \ref{fig:scatter_pb} show the joint
distributions for a combination of the angular number counts of halos,
the lensing power spectrum, or the lensing bispectrum of equilateral 
triangle
configurations. 
The halo model fairly well reproduces the joint distributions over
a range of multipoles (the width and the direction of the
cross-correlation). However, we note that the agreement for the
bispectrum is not relatively as good as for the power spectrum,
reflecting the limitation of the multivariate Gaussian assumption for the
bispectrum distribution.

\begin{figure}
\centering
\includegraphics[width=3.6in]{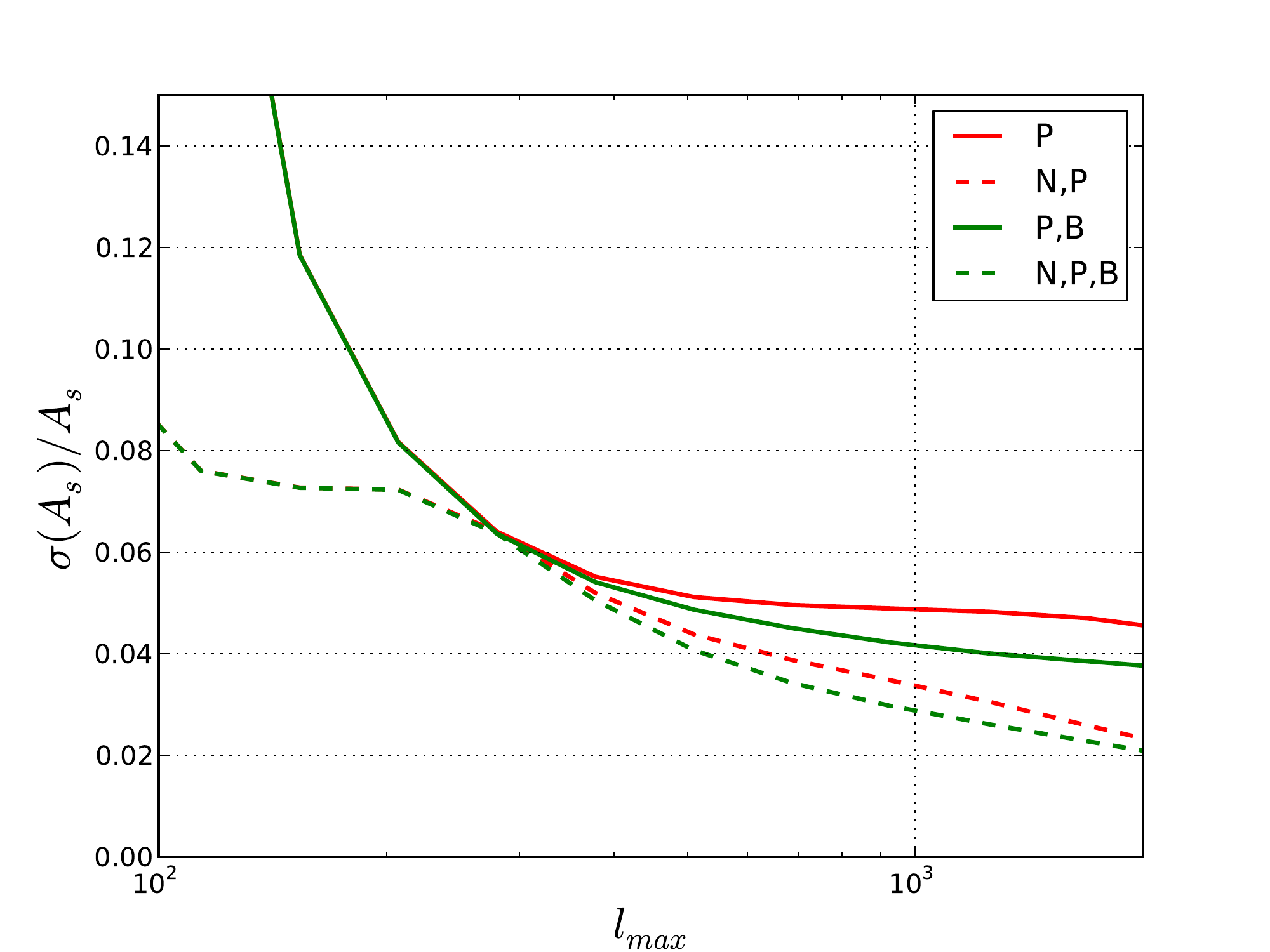}
\caption{
Forecast error on the amplitude of scalar perturbations $A_S$ as a
 function of the maximum observed multipole $l_\text{max}$, when
 marginalizing over $\Omega_m$, and keeping all other parameters fixed. 
We here assumed a hypothetical survey that is characterized by
area $15 000$ sq. deg.
and the redshift distribution following \cite{OguriTakada:11} with
 $\ave{z_s} = 3 z_0 = 1$. 
We included the galaxy shape noise
 contamination, 
assuming
$\sigma_\epsilon = 0.22$ and $n_\text{gal} = 20
 \text{arcmin}^{-2}$ for the number density and the rms intrinsic
 elllipticities.
The solid red curve shows the error when including the power spectrum
 information alone. The solid-green and dashed curves show the results
 when further including the bispectrum measurements of equilateral
 triangle configuration and the cluster number counts.
}
\label{fig:fisher_as}
\end{figure}
Having found that the halo model fairly well describes the joint
likelihood functions between the cluster number counts and the lensing
correlation functions, we now discuss how a future survey can improve
cosmological constraints based on the joint measurements of the
different observables obtained from the same survey data. As one
demonstration, we consider only two cosmological parameters, the matter
density parameter $\Omega_{\rm m}$ and the amplitude parameter of the
primordial curvature perturbation $A_s$, both of which are sensitive to
the amplitudes of the number counts and the lensing correlation
functions and therefore are most affected by the HSV effect. However,
note that we fix all other parameters to their fiducial values.  Assuming
the multivariate Gaussian likelihood, we use the Fisher information
matrix formalism to perform a parameter forecast:
\begin{equation}
F_{ij} =  \frac{\partial \mbf{D}^t}{\partial p_i} \mbf{\Sigma}^{-1}  
\frac{\partial \mbf{D}}{\partial p_j},
\end{equation}
where $p_i$ denotes the $i$-th cosmological parameter
($\Omega_{\rm m}$ or $A_s$). To make parameter forecast for a
hypothetical future survey, we include the shape noise contamination to
the lensing power spectrum and bispectrum covariances for which we model
assuming $\bar{n}_g=20~$arcmin$^{-2}$ and $\sigma_\epsilon=0.22$ for the
mean number density of source galaxies and the rms intrinsic
ellipticities, respectively. We assume $\Omega_S=15,000$~sq. degrees for
survey area, and model the redshift distribution of source galaxies by
an analytical model that is given by $\ave{z_s}=3z_0=1$ in Eq.~(17) in
Oguri \& Takada \cite{OguriTakada:11}. These survey parameters resemble
those expected for a Euclid-type survey.

Fig.~\ref{fig:fisher_as} shows the expected precision of $A_s$ including
marginalization over $\Omega_{\rm m}$, as a function of maximum
multipole up to which we include the power spectrum information (and
also the lensing bispectrum). Note that, for the bispectrum, we
included all the equilateral 
triangle configurations available
over a range of multipoles up to a given maximum multipole, but did not
include other triangle information. 
The figure shows that combining the
cluster number counts with the power spectrum and bispectrum measurements
for $l_{\rm max}=1000$, which is the target maximum multipole for the
Euclid survey, tightens the error by a factor of $30$--$40\%$. This
improvement is equivalent to a factor 2 wider survey area.

\section{Conclusion}

In this paper, we presented a simple and general formalism to compute
the halo sample variances for the cluster counts, and any $n$-point
function for the matter density or any projected density field, such as
cosmic shear or the thermal Sunyaev-Zel'dovich effect. These results
rely only on the assumptions built into the halo model, provide a good
fit to the simulation from Ref.~\cite{Satoetal:09}, and allow for an
intuitive understanding of all the terms.

We presented a simple ansatz for the joint likelihood of cluster counts,
power spectrum and bispectrum of the lensing convergence, and showed its
relatively good agreement with simulation. We used this joint likelihood
to estimate that constraints on cosmological parameters such as
$\Omega_m$ and $A_s$ can be improved by $30\%$ ($40\%$) if
one combines the cluster counts with the power spectrum measurement 
(further combined with the lensing
bispectrum).
This is equivalent to doubling
the survey volume.

Taking into account the specific geometry of the survey (and not only
its volume), as well as the selection functions for the cluster counts
and the uncertainty on the mass determination, constitute interesting
extensions of this study which we leave for future work.


\smallskip{\em Acknowledgments.--} We thank Issha Kayo for providing us
with his analyzed data from the simulations in \cite{Satoetal:09}, and
for his help and kindness throughout this project. ES would like to
thank Elisabeth Krause for helpful discussion, Simone Ferraro for his
many useful comments throughout this project as well as reading an early
version of this paper, and the Kavli IPMU for their hospitality.
DNS and ES acknowledge support from NSF Grant AST-1311756,  NASA Grant
11-ATP-090, NASA ROSES grant 12-EUCLID12-0004 and the Euclid Consortium.
MT was supported by World
 Premier International Research Center Initiative (WPI Initiative),
 MEXT, Japan, by the FIRST program ``Subaru Measurements of Images and
 Redshifts (SuMIRe)'', CSTP, Japan, and by Grant-in-Aid for
 Scientific Research from the JSPS Promotion of Science (No.~23340061 and 26610058).

\bibliography{final_version}

\appendix

\section{Matter $N$-point functions: expectation values and covariances}
\label{app:pp} 

\subsection{Expectation values: halo decomposition}

In this subsection we derive the expectation value of our power spectrum estimator without using the general results from Section~\ref{subsec:general_derivation}, in order to see in more details why the expectation value is not affected by the finite size effect. The same reasoning applies to any $n$-point function as we shall explain.

Our estimator for the power spectrum is given by Eq~\eqref{eq:p_estimator}, where the matter overdensity is described in the halo model by Eq~\eqref{eq:def_hdelta}. The expectation value for the power spectrum estimator is obtained directly from:
\beq
\bal
\ave{ \hdelta(\vq_1) \hdelta(\vq_2) }
&=
\ave{ \sum_{i,j} \left( \frac{m_i m_j}{\bar{\rho}^2} \right) 
u_i(\vq_1)u_j(\vq_2) \,\,
\hn_i(\vq_1, \delta_b) \hn_j(\vq_2, \delta_b) } \\
&=
\sum_{i,j} \left( \frac{m_i m_j}{\bar{\rho}^2} \right) 
u_i(\vq_1)u_j(\vq_2) \,\,
\ave{ \hn_i(\vq_1, \delta_b) \hn_j(\vq_2, \delta_b) } . \\
\eal
\eeq 
Thus we only need to compute the quantity $\ave{ \hn_i(\vq_1, \delta_b) \hn_j(\vq_2, \delta_b) }$. We decompose the averaging procedure into marginalizing over the Poisson sampling, then the underlying density $\rho_\text{lin}$, and eventually the local average overdensity $\delta_b$. The First average is obtained from Eq~\eqref{eq:poisson}, and gives the usual Poisson shot noise:
\beq
\bal
\ave{ \hn_i(\vx_1) \hn_j(\vx_2) }_\text{Pois.}
&=
n_i(\vx_1) n_j(\vx_2)
+ \deltak_{i,j} \deltad(\vx_1 -\vx_2) n_i(\vx),
\eal
\eeq
where again, we defined $n_i(\vx) \equiv \ave{\hn_i(\vx)}_\text{Pois.}$. After Fourier transform, this becomes:
\beq
\bal
\ave{ \hn_i(\vq_1) \hn_j(\vq_2) }_\text{Pois.}
&=
n_i(\vq_1) n_j(\vq_2) 
+ \deltak_{i,j}  n_i(\vq_1+\vq_2) .\\
\eal
\eeq
Averaging over the underlying density field at fixed $\delta_b$ gives:
\beq
\bal
\ave{ \hn_i(\vq_1) \hn_j(\vq_2) }_{\text{Pois.}, \rholin | \delta_b}
&=
\ave{ n_i(\vq_1) n_j(\vq_2) }_{ \rholin | \delta_b }
+ \deltak_{i,j}  \on_i(\delta_b) \left( 2\pi \right)^3 \deltad(\vq_1+\vq_2) \\
&=
\on_i(\delta_b) \on_j(\delta_b)\ave{ \delta^h_i(\vq_1) \delta^h_j(\vq_2) }_{ \rholin | \delta_b }
+ \deltak_{i,j}  \on_i(\delta_b) \left( 2\pi \right)^3 \deltad(\vq_1+\vq_2) \\
&=
\left[ \on_i(\delta_b) \on_j(\delta_b)   b_ib_j P_\text{lin} (q_1) 
+ \deltak_{i,j}  \on_i(\delta_b) \right] \left( 2\pi \right)^3 \deltad(\vq_1+\vq_2) .\\
\eal
\eeq
Here, we introduced the usual halo number overdensity $\delta_i^h$ for halos of mass $m_i$, and used the linear bias to write $\ave{\delta_i^h \delta_j^h} = \left( 2\pi \right)^3 \deltad b_i b_j P_\text{lin}$.
Switching to discrete Fourier transform, this becomes:
\beq
\bal
&\ave{ \hn_i(\vq_1) \hn_j(\vq_2) }_{\text{Pois.}, \rholin | \delta_b}
=
\left[ \on_i(\delta_b) \on_j(\delta_b)   b_ib_j P_\text{lin} (q_1) 
+ \deltak_{i,j}  \on_i(\delta_b) \right] V_S \deltak_{\vq_1+\vq_2} .\\
\eal
\eeq
Hence the expectation value of the power spectrum at fixed $\delta_b$:
\beq
\bal
\ave{ \hat{P}(k) }_{| \delta_b} 
&= 
\sum_{i,j}  \left( \frac{m_i m_j}{\bar{\rho}^2} \right) 
u_i(k)u_j(k) \,\,
\on_i(\delta_b) \on_j(\delta_b)   b_ib_j P_\text{lin} (k) 
\,\, + \,\,  \sum_i \left( \frac{m_i}{\bar{\rho}} \right)^2
|u_i(k)|^2\,\,
 \on_i(\delta_b) .
 \eal
\eeq
This is nothing but the halo decomposition $P=P^{1h} + P^{2h}$, except for the extra $\delta_b$-dependence. To marginalize over $\delta_b$, we simply use Eq~\eqref{eq:linear_bias}:
\beq
\bal
\ave{ \hat{P}(k) }
&=
\sum_{i,j}  \left( \frac{m_i m_j}{\bar{\rho}^2} \right) 
u_i(k)u_j(k) \,\,
\on_i \on_j \left[  1 + \sigma_m^2 b_ib_j  \right]   b_ib_j P_\text{lin} (k)
\,\, + \,\,  \sum_i \left( \frac{m_i}{\bar{\rho}} \right)^2
|u_i(k)|^2\,\,
 \on_i .
 \eal
\eeq
This averaging procedure is equivalent to the one described by eq~\eqref{eq:substitution_n} in Section~\ref{subsec:general_derivation}. Provided that $\sigma_m^2 b_ib_j \ll 1$, we see that the finite size of the box does not lead to any significant bias in our power spectrum estimator.

The same reasoning applies for any $n$-point function: one can disregard the effect of $\delta_b$ when computing expectation values, and using the Poisson property Eq~\eqref{eq:poisson} leads to the standard halo decomposition $P_N = P_N^{1h} + ... + P_N^{Nh}$.

\subsection{Covariances}

In this subsection, we derive and discuss the general result Eq~\eqref{eq:simplified_exp}. We start with observables $\hat{f}$ and $\hat{g}$ that are determined by the halo counts $\left\{ \hn_i (\vx) \right\}$, and we call $\bar{f}(\delta_b) = \ave{\hat{f}}_{\delta_b}$ (and similarly for $\hat{g}$). We Taylor expand $\bar{f}$ (and $\bar{g}$) around $\delta_b=0$:
\beq
\bar{f}(\delta_b) = \bar{f}(0) 
+ \left.\frac{\partial \bar{f}}{\partial \delta_b}\right|_{\delta_b=0} \delta_b 
+ \frac{1}{2} \left.\frac{\partial^2 \bar{f}}{\partial \delta_b^2}\right|_{\delta_b=0} \delta_b^2 
+ \mathcal{O}(\delta_b^3) ,
\eeq
marginalizing over $\delta_b$ then yields:
\beq
\ave{\bar{f}(\delta_b)} = \bar{f}(0) 
+ \frac{1}{2} \left.\frac{\partial^2 \bar{f}}{\partial \delta_b^2}\right|_{\delta_b=0} \sigma_m^2 
+ \mathcal{O}(\sigma_m^4) ,
\eeq
and:
\beq
\ave{ \bar{f}(\delta_b) \bar{g}(\delta_b) } 
= \ave{ \bar{f}(\delta_b)} \ave{\bar{g}(\delta_b) } 
+  \left.\frac{\partial \bar{f}}{\partial \delta_b}\right|_{\delta_b=0}  \left.\frac{\partial \bar{g}}{\partial \delta_b}\right|_{\delta_b=0}\sigma_m^2 
+ \mathcal{O}(\sigma_m^4) .
\eeq
Deriving relation~\eqref{eq:general_exp} is now straightforward:
\beq
\bal
\text{Cov}\left[ \hat{f}, \hat{g} \right]
&\equiv \ave{\ave{ \hat{f} \hat{g} }_{| \delta_b} }_{\delta_b} 
- \ave{\ave{ \hat{f} }_{| \delta_b} }_{\delta_b} \ave{\ave{ \hat{g} }_{| \delta_b} }_{\delta_b} \\
&= \ave{\ave{ \hat{f} \hat{g} }_{| \delta_b} }_{\delta_b} 
- \ave{ \bar{f}(\delta_b)}_{\delta_b} \ave{\bar{g}(\delta_b) }_{\delta_b} \\
&= \ave{\ave{ \hat{f} \hat{g} }_{| \delta_b} }_{\delta_b}
- \ave{ \bar{f}(\delta_b) \bar{g}(\delta_b) }_{\delta_b}
+ \left.\frac{\partial \bar{f}}{\partial \delta_b}\right|_{\delta_b=0} \left.\frac{\partial \bar{g}}{\partial \delta_b}\right|_{\delta_b=0} \sigma_m^2 
+ \mathcal{O}(\sigma_m^4) \\
&= \ave{\ave{ \hat{f} \hat{g} }_{| \delta_b} 
- \bar{f}(\delta_b) \bar{g}(\delta_b) }_{\delta_b}
+ \left.\frac{\partial \bar{f}}{\partial \delta_b}\right|_{\delta_b=0} \left.\frac{\partial \bar{g}}{\partial \delta_b}\right|_{\delta_b=0} \sigma_m^2 
+ \mathcal{O}(\sigma_m^4) \\
&= \ave{ \text{Cov}\left[ \hat{f}, \hat{g} \right]_{| \delta_b} }_{\delta_b}
+ \left.\frac{\partial \bar{f}}{\partial \delta_b}\right|_{\delta_b=0} \left.\frac{\partial \bar{g}}{\partial \delta_b}\right|_{\delta_b=0} \sigma_m^2 
+ \mathcal{O}(\sigma_m^4) . \\
\eal
\eeq
And as we've seen, marginalizing over $\delta_b$ reduces to the substitution~\eqref{eq:substitution_n}, which leads to negligible terms (in the same exact way as in the example of the power spectrum expectation value above). Hence we get Eq~\eqref{eq:simplified_exp}:
\beq
\text{Cov}\left[  \hat{f}, \hat{g}  \right]
\simeq
\text{Cov}\left[  \hat{f}, \hat{g}  \right]_{\text{Pois.}, \rho_\text{lin} | \delta_b=0} 
+ \sigma_m^2  \left.\frac{\partial \bar{f}}{\partial \delta_b}\right|_{\delta_b=0} \left.\frac{\partial \bar{g}}{\partial \delta_b}\right|_{\delta_b=0} .\\
\eeq
The second term is the halo sample variance, and is readily computed from the halo decomposition of the $N$-point functions and linear biasing. The first term is the standard covariance. It is computed by using the expression of our estimator as a product of $\hdelta$, and the decomposition of correlation functions into connected correlation functions \cite{Bernardeauetal:02}:
\beq
\bal
\ave{\hdelta(\vx_1) ... \hdelta(\vx_n)}
&=
\ave{\hdelta(\vx_1), ... ,\hdelta(\vx_n)}_c
+ 
\sum_S
\prod_{s_i \in S}
\ave{\hdelta_{s_i(1)}, ..., \hdelta_{s_i(\# s_i)}}_c ,
\eal
\eeq
where the sum is over all the proper partitions of $\{ 1, ..., n \}$. This yields the following standard covariance for the power spectrum:
\beq
\bal
\text{Cov}\left[  \hat{P}(k), \hat{P}(k')  \right]_{| \delta_b = 0}
&=
\frac{2 \delta^K_{k,k'} } {N(k)}   \bar{P}^2(k)
+\frac{1}{V_S} \bar{T}(\vec{q}, -\vec{q}, \vec{q}^{\, \prime}, -\vec{q}^{\, \prime})^{\| \vq \| \simeq k, \| \vq ' \| \simeq k'} .\\
\eal
\eeq
This procedure also yields the results of \cite{Kayoetal:12} for the bispectrum covariance:
\beq
\label{eq:std_covb}
\bal
&\text{Cov} \left[  \hat{B}_{(k_1, k_2, k_3)} , \hat{B}_{(k_1', k_2', k_3')}  \right] _{| \delta_b = 0} 
=
V_S 
\frac{  \deltak_{k_1,k_1'} \deltak_{k_2,k_2'} \deltak_{k_3, k_3'}  }{N_\Delta (k_1, k_2, k_3)}
\, \bar{P}_{(k_1)} \bar{P}_{( k_2)} \bar{P}_{( k_3)}
+ \text{5 permutations of the } k_{i'} \\
&\hspace{1.5cm}+
\frac{\deltak_{k_1', k_3}}{N_\Delta (k_1, k_2, k_3) N_\Delta (k_1', k_2', k_3')}
\sum_{1, 2, 3 \atop 2', 3'}
\deltak_{1+2+3} \deltak_{3+2'+3'} 
\, \bar{B}_{(1, 2, 3)} \, \bar{B}_{(3, 2', 3')}
+ \text{8 permutations} \\
&\hspace{1.5cm}+
\frac{\deltak_{k_1, k_1'}}{N_\Delta (k_1, k_2, k_3) N_\Delta (k_1', k_2', k_3')}
\sum_{1, 2, 3 \atop 2', 3'}
\deltak_{1+2+3} \deltak_{-1+2'+3'}
\, \bar{P}_{(1)} \, \bar{T}_{(2, 2', 3, 3')} 
+ \text{8 different choices } (i,j') \\
&\hspace{1.5cm}+
\frac{1}{V_S}
\frac{1}{N_\Delta (1, 2, 3) N_\Delta (1', 2', 3')}
\sum_{1, 2, 3 \atop 1', 2', 3'}
\deltak_{1+2+3} \deltak_{1'+2'+3'} 
\bar{P}_{6} (1, 2, 3, 1', 2', 3') .
\eal
\eeq
As well as their result for the cross-covariance between power spectrum and bispectrum:
\beq
\label{eq:std_covpb}
\bal
\text{Cov} \left[  \hat{P}_{(k)}, \hat{B}_{(k_1, k_2, k_3)}  \right] _{| \delta_b = 0} 
&= \frac{2 \deltak_{k, k_1}}{N(k_1)} \bar{P}(k_1) \bar{B}(k_1, k_2, k_3) + \text{2 perm.} \\
&+ \frac{1}{V_S} \int \frac{d\psi}{2\pi} \bar{P}_5(k, -k, k_1, k_2, k_3) ,\\
\eal
\eeq
where $\psi$ is the angle between $\vk$ and $\vk_1$.

\end{document}